\documentclass[aps,nofootinbib,notitlepage,showpacs,longbibliography]{revtex4-1}

\usepackage[CJKbookmarks, pdftex, bookmarksnumbered, bookmarksopen, colorlinks, citecolor=magenta, linkcolor=magenta]{hyperref}

\usepackage{amstext,amsmath,amssymb,amsfonts,bbm}
\usepackage[latin1]{inputenc}
\usepackage{graphicx}
\usepackage{epstopdf}
\usepackage{amsthm}
\usepackage{tocvsec2}
\usepackage{enumerate}
\usepackage{subfigure}
\usepackage{color}

\usepackage{multirow} \usepackage{rotating}

\def\beq{\begin{equation}}
\def\be{\begin{equation}}
\def\ee{\end{equation}}
\def\bes{\begin{eqnarray}}
\def\ees{\end{eqnarray}}

\begin{document}

\title{(3+1)-Formulation for Gravity with Torsion and Non-Metricity: The
Stress-Energy-Momentum Equation}

\author{Seramika Ariwahjoedi$^{1,2}$, Agus Suroso$^{2,3}$, Freddy P.
Zen$^{2,3}$\vspace{1mm}
 }

\affiliation{$^{1}$Research Center for Physics, Indonesian Institute of Sciences (LIPI), Serpong PUSPIPTEK Area, Tangerang 15310, Indonesia.\\
 $^{2}$Theoretical Physics Laboratory, THEPI Division, Institut Teknologi
Bandung, Jl. Ganesha 10 Bandung 40132, West Java, Indonesia.\\
 $^{3}$Indonesia Center for Theoretical and Mathematical Physics
(ICTMP), Indonesia.}

\begin{abstract}
\noindent We derive the generalized Gauss-Codazzi-Mainardi (GCM) equation
for a general affine connection with torsion and non-metricity. Moreover,
we show that the metric compatibility and torsionless condition of
a connection on a manifold are inherited to the connection of its
hypersurface. As a physical application to these results, we derive
the (3+1)-Einstein Field Equation (EFE) for a special case of Metric-Affine
$f\left(\mathcal{R}\right)$-gravity when $f\left(\mathcal{R}\right)=\mathcal{R}$,
the Metric-Affine General Relativity (MAGR). Motivated by the concept
of geometrodynamics, we introduce additional variables on the hypersurface
as a consequence of non-vanishing torsion and non-metricity. With
these additional variables, we show that for MAGR, the energy, momentum,
and the stress-energy part of the EFE are dynamical, i.e., all of
them contain the derivative of a quantity with respect to the time
coordinate. For the Levi-Civita connection, one could recover the
Hamiltonian and the momentum (diffeomorphism) constraint, and obtain
the standard dynamics of GR.
\end{abstract}
\maketitle

\section{Introduction}

With the recent discoveries of gravitational waves and black holes,
General Relativity (GR) has been one of the most successful theories
in physics. However, several problems have not been covered by GR,
these include dark matter, dark energy, and quantum gravity. Attempts
to solve these problems have led to some motivation in modifying the
original  General Relativity, proposed by Einstein in 1915. Different
types of modification had flourished with rapid progression; for a
complete review on theories of modified gravity, one could consult
\cite{Oikonomou}. A specific way to achieve a modified theory is
by adding higher-order terms on curvature to the Einstein-Hilbert
action. In $f(\mathcal{R})$-gravity, the higher-order terms are restricted
to the power series of the Ricci scalar curvature \cite{Oikonomou,Sotiriou}.
Another way to modify GR is by considering a general affine connection
with torsion, and even non-metricity, hence the connection is not
necessarily Levi-Civita as in the standard GR.

Different theories of gravity do not use curvature to describe the
effect of gravity, they instead use torsion in Teleparallel Gravity
\cite{Pereira}, or non-metricity in Symmetric Teleparallel Gravity
\cite{Nester,Jimenez}. There exists even a more general theory that
uses both torsion and non-metricity (but zero curvature) in General
Teleparallel Quadratic Gravity \cite{Jimenez3}. On the other hand,
the Metric-Affine-Gravity (MAG) is a curvature theory of gravity that
includes both non-metricity and torsion for the connection \cite{Hehl,Iosifidis}.
The choice of action for MAG could vary greatly, for example, the
Ricci scalar $\mathcal{R}$, the power series of Ricci scalar $f\left(\mathcal{R}\right)$,
and other exotic actions as discussed in \cite{Iosifidis}. In this
article, we only consider Metric-Affine General Relativity (MAGR),
a class of MAG with Ricci scalar $\mathcal{R}$ as the curvature part
of action. It could also be considered as a special case of the most
general version of the $f(\mathcal{R})$-gravity, the Metric-Affine
$f(\mathcal{R})$-gravity, that is a modified theory of gravity with
$f(\mathcal{R})$-action and general affine connection \cite{Sotiriou}.
The existence of torsion in the connection is important for the coupling
of gravity with fermions \cite{Hehl2}, while the non-metricity part
is important if one considers certain kinds of matter fields that
contain a non-vanishing symmetric part of the hypermomentum \cite{Sotiriou,Hehl3,Hehl4,Hehl5}.

To obtain the dynamics of a covariant gravitational theory, one needs
to decompose the covariant fields into their temporal and spatial
parts via the (3+1) decomposition. It was first  considered by Darmois
in \cite{Hehl}, Lichnerowicz in \cite{Lichnerowicz1,Lichnerowicz2,Lichnerowicz3},
then by Choquet-Bruhat in \cite{Choquet1,Choquet2}. However, it gained
a lot of attention after the work by Arnowitt, Deser, and Misner in
\cite{ADM}. The (3+1) ADM decomposition is a formulation of GR which
split the spacetime and the covariant fields into the temporal and
spatial components, thus allowing one to obtain the dynamics of the
variables on the 3D hypersurface via the equation of motions. At the
core of the ADM formulation, lies the Gauss-Codazzi-Mainardi equations,
which relates the intrinsic (Riemann) curvatures of a manifold with
the extrinsic and intrinsic curvature of its submanifold. The result
of the formulation is the (3+1) Einstein Field Equations (EFE): 4
constraint equations and 6 dynamical ones.

Attempts to apply the (3+1) formulation to $f(\mathcal{R})$-gravity
had been done for special cases; in $f(\mathcal{R})$-gravity with
Levi-Civita connection \cite{Nathalie,Paschalidis} and in teleparallel-gravity
with torsion \cite{Shapiro}. Even more, the Hamiltonian analysis
and possible quantization procedure of $f(\mathcal{R})$-gravity had
been done in \cite{Olmo2,Ma1,Ma2}. Little attention had been given
to the (3+1) formulation of gravity with non-metricity, with the latest
work is concerning the derivation of the (1+3) Raychaudhuri equation
\cite{Iosifidis,Iosifidis2}. However, for various reasons, it is
interesting to perform the (3+1) decomposition for MAG and $f(\mathcal{R})$-gravity
which includes non-metricity, particularly, if one considers matter
fields with a general hypermomentum. With this motivation in mind,
in this article, we apply the (3+1)  ADM formulation for Metric-Affine
General Relativity (MAGR) with general affine connection that includes
torsion and non-metricity, as a special case to MAG and $f(\mathcal{R})$-gravity.

The article is organized as follows: In Section II, we derive the
generalized Gauss-Codazzi-Mainardi (GCM) equations for a general affine
connection with torsion and non-metricity. A similar treatment is
also applied to torsion, resulting in a torsion (3+1) decomposition.
These results are valid for manifold with any dimension. For the derivation,
we use 2 different definitions of extrinsic curvature that exist in
the literature; both differ by the quantity $\nabla_{X}g$ (with $\left(\nabla,g\right)$
are the connection and the metric on the manifold, and $X$ is a vector
on the hypersurface) if the connection is non-metric. The generalized
GCM equations are written in terms of these quantities, together with
the definition of the acceleration tensor.

In Section III, we show that the metric compatibility and the torsionless
condition of a connection $\nabla$ on a manifold are inherited to
the connection $\,^{3}\nabla$ on the hypersurface. This was already
a well-known result and was used to derive the original GCM equation
for the Levi-Civita connection. Since the generalized Riemann curvature
tensor of an affine connection is lacking the symmetries possessed
by the Levi-Civita connection, we need to define three different types
of independent contraction of the Riemann tensor, following the definition
introduced in \cite{Iosifidis3,Jimenez2,Iosifidis4}. These contractions
are important for the derivation of the Einstein tensor in the next
section.

Section IV consists of the physical application of the generalized
GCM and the torsion decomposition. Minimization of the $f(\mathcal{R})$-action
with respect to the variation of metric, connection, and a Lagrange
multiplier results in 2 Euler-Lagrange equations and 1 constraint
equation. However, considering the scope of this article, our focus
will be on the generalized stress-energy-momentum equation (EFE):
the one that comes from the variation of the metric. We perform the
ADM formulation to the generalized EFE using the generalized GCM equation
we derive in Section II, resulting in 3 parts of the equation: the
energy, momentum, and stress-energy equation. With the introduction
of some additional variables on the hypersurface, we could write the
energy, momentum, and stress-energy part of the EFE in terms of tensor
fields defined only on the hypersurface, motivated by the concept
of geometrodynamics in \cite{Wheeler}. An interesting result here
is, unlike the standard GR case, there is no constraint originating
from the generalized EFE; all the 3 equations contain dynamics, i.e,
contain the derivative of a quantity with respect to the time coordinate.
However, with the Levi-Civita condition, one could recover the Hamiltonian
and momentum (or diffeomorphism) constraint, together with the standard
dynamics of GR.

We discuss some subtleties on the results in Subsection VA, and for
the completeness of the analysis in this article, we present a special
case of (3+1) decomposition of the 2 remaining equations of motion:
the hypermomentum and the projective constraint equation. This is
done in Subsection VB. The detailed derivation and general treatment
of these equations will be presented in our companion article. Finally,
we conclude our works in Subsection VC.

\section{The Generalized Gauss-Codazzi-Mainardi Equations}

\subsubsection*{The Metric and Connection on the Hypersurface}

Let $\mathcal{M}$ be an $n$-dimensional manifold, equipped with
a (pseudo)-Riemannian metric $g$ and an affine connection $\nabla.$
In the most general case, $g$ and $\nabla$ are independent of one
another. Let $\Sigma$ be the hypersurface, namely, an ($n$-1)-dimensional
submanifold embedded on $\mathcal{M}.$ One could define the unit
vector normal to $\Sigma$ at each point $p\in\Sigma$, namely, $\hat{n}_{p}$,
satisfying $g_{p}\left(\hat{n}_{p},\hat{n}_{p}\right)=\pm1,$ where
the $-$ sign (the one on the upper side) and the $+$ sign (the one
on the lower side) depends on whether $\left(M,g\right)$ is Riemannian
or Lorentzian, respectively. From this point up to the rest of the
article, the signature on the upper side of the equation is the one
valid for the Riemannian case, while the lower side is for the Lorentzian's.
The existence of metric $g$ on $\mathcal{M}$ induces a metric on
the hypersurface $\Sigma$ as follows:
\begin{equation}
\,^{3}q=g\mp\hat{n}^{*}\otimes\hat{n}^{*},\label{eq:1a}
\end{equation}
with $\hat{n}^{*}\in T_{p}^{*}\mathcal{M}$ is the covariant vector
to $\hat{n},$ satisfying $\hat{n}^{*}=g\left(\hat{n},\cdot\right)$
(here, we omit the label $p$ for simplicity).

Let $V,X\in T_{p}\Sigma$, then one could construct a new vector $\nabla_{V}X$
$\in T_{p}\mathcal{M},$ not necessarily an element of $T_{p}\Sigma$,
since $\nabla$ is defined on $\mathcal{M}$. The decomposition of
$\nabla_{V}X$ into the components parallel and perpendicular to $\Sigma$
is the following:
\begin{equation}
\nabla_{V}X=\underset{\left(\nabla_{V}X\right)_{\parallel}}{\underbrace{\nabla_{V}X\mp g\left(\nabla_{V}X,\hat{n}\right)\hat{n}}}\pm\underset{\left(\nabla_{V}X\right)_{\perp}}{\underbrace{g\left(\nabla_{V}X,\hat{n}\right)\hat{n}}.}\label{eq:x}
\end{equation}
The labels $\parallel$ and $\perp$ denote, respectively, the parallel
part of $\nabla_{V}X$ which lies on $\Sigma$ and the part normal
to $\Sigma.$ The covariant derivative $\nabla$ on the quantity $\left(\nabla_{V}X\right)_{\parallel}$
associates $V,X\in T_{p}\Sigma$ to a transformed vector $\left(\nabla_{V}X\right)_{\parallel}\in T_{p}\Sigma$,
hence, one defines:
\begin{equation}
\left(\nabla_{V}X\right)_{\parallel}:=\,^{3}\nabla_{V}X=\nabla_{V}X\mp g\left(\nabla_{V}X,\hat{n}\right)\hat{n},\label{eq:2}
\end{equation}
where $\,^{3}\nabla$ is the covariant derivative on $\Sigma$. It
could be easily shown that $\,^{3}\nabla$ is an affine connection.
One could define an entirely different affine connection $^{3}\nabla$
on $\Sigma$, which in general, is not induced from $\nabla$. However,
since the objective of the work is to obtain the generalized Gauss-Codazzi
relation, we only consider the case where $^{3}\nabla$ is induced
by $\nabla$ as in (\ref{eq:x}).

\subsubsection*{Extrinsic Curvature of the First Kind}

The component of the perpendicular part of (\ref{eq:x}) is defined
as the extrinsic curvature of $\Sigma$ in the direction of $\left(V,X\right)$
as follows \cite{Baez}:
\begin{equation}
\left(\nabla_{V}X\right)_{\perp}:=K\left(V,X\right)=\pm g\left(\nabla_{V}X,\hat{n}\right).\label{eq:3}
\end{equation}
Written in terms of components, one could show that the extrinsic
curvature of the first kind satisfies: 
\begin{equation}
K\left(V,X\right)=\pm V^{\mu}X^{\beta}\omega_{\mu\,\:\,\beta}^{\,\:\,\alpha}n_{\alpha},\label{eq:z}
\end{equation}
using the fact that $n_{\alpha}V\left(X^{\alpha}\right)=0$, since
$X,V\in T_{p}\Sigma$ are perpendicular to $\hat{n}$. The quantity
$\omega_{\mu\,\:\,\beta}^{\,\:\,\alpha}\partial_{\alpha}=\nabla_{\mu}\partial_{\beta}$
is the vector potential of $\nabla$. Notice that in general, $\omega_{\mu\,\:\,\beta}^{\,\:\,\alpha}$
of an affine connection does not possess any symmetry in the indices
$\left(\mu,\alpha,\beta\right),$ therefore, one needs to be careful
not to switch the position of these indices. With definitions (\ref{eq:2})
and (\ref{eq:3}), (\ref{eq:x}) could be written as:
\begin{equation}
\nabla_{V}X=\,^{3}\nabla_{V}X+K\left(V,X\right)\hat{n}.\label{eq:y}
\end{equation}
There exists another definition of extrinsic curvature that will be
explored in the next sections.

\subsubsection*{The Torsion Tensor}

The torsion tensor on $\mathcal{M}$ is defined as:
\begin{equation}
T\left(X,Y\right)=\nabla_{X}Y-\nabla_{Y}X-\left[X,Y\right],\qquad X,Y\in T_{p}\mathcal{M}.\label{eq:torsion}
\end{equation}
For the case where $X,Y\in T_{p}\Sigma,$ using the parallel and perpendicular
decomposition of $\nabla_{V}X$ as in (\ref{eq:y}), one could obtain:
\begin{equation}
T\left(X,Y\right)=\,^{3}T\left(X,Y\right)+\left(K\left(X,Y\right)-K\left(Y,X\right)\right)\hat{n},\label{eq:torsion-1}
\end{equation}
with:
\[
\,^{3}T\left(X,Y\right)=\,^{3}\nabla_{X}Y-\,^{3}\nabla_{Y}X-\left[X,Y\right],
\]
is the torsion on the hypersurface $\Sigma$. Moreover, using (\ref{eq:z}),
one could obtain:
\[
K\left(X,Y\right)-K\left(Y,X\right)=\pm g\left(T\left(X,Y\right),\hat{n}\right),
\]
and hence (\ref{eq:torsion-1}) could be regarded as the torsion decomposition
into the parallel and normal part of the hypersurface. The torsion
tensor is antisymmetric in the first and second argument:
\[
T\left(X,Y\right)=-T\left(Y,X\right),\qquad\forall\;X,T\,\in T_{p}\mathcal{M}.
\]

\subsubsection*{The Generalized (Riemann) Intrinsic Curvature}

The original Riemann curvature is defined only for the curvature of
a Levi-Civita connection in the context of Riemannian geometry. In
this article, we use a similar definition to describe the intrinsic
curvature of a general affine connection in non-Riemannian geometry:
\begin{equation}
\boldsymbol{R}\left(U,V\right)X=\nabla_{U}\nabla_{V}X-\nabla_{V}\nabla_{U}X-\nabla_{\left[U,V\right]}X,\qquad U,V,X\in T_{p}\mathcal{M}.\label{eq:Riemann}
\end{equation}
To obtain the intrinsic curvature on $\Sigma$, let $U,V,X\in T_{p}\Sigma$,
then with the definitions in (\ref{eq:2}) and (\ref{eq:3}), one
could obtain the double derivative of $X$:
\begin{equation}
\nabla_{U}\nabla_{V}X=\,^{3}\nabla_{U}\,^{3}\nabla_{V}X+K\left(V,X\right)\nabla_{U}\hat{n}+\left(U\left[K\left(V,X\right)\right]+K\left(U,\,^{3}\nabla_{V}X\right)\right)\hat{n},\label{eq:4}
\end{equation}
from a direct calculation. Notice here that $U\left[f\left(x\right)\right]$
denotes a vector $U$ acting on a scalar function $f\left(x\right)$
as $U^{\mu}\partial_{\mu}f\left(x\right)$. Moreover one could have
the following quantity:
\begin{align}
\nabla_{V}\nabla_{U}X & =\,^{3}\nabla_{V}\,^{3}\nabla_{U}X+K\left(U,X\right)\nabla_{V}\hat{n}+\left(V\left[K\left(U,X\right)\right]+K\left(V,\,^{3}\nabla_{U}X\right)\right)\hat{n},\label{eq:5}\\
\nabla_{\left[U,V\right]}X & =\,^{3}\nabla_{\left[U,V\right]}X+K\left(\left[U,V\right],X\right)\hat{n},\label{eq:6}
\end{align}
using the fact that $\left[U,V\right]\in T_{p}\Sigma$ for $U,V\in T_{p}\Sigma$.
By a little tensor algebra, one could show that the following relation
is valid:
\begin{align}
K\left(U,\,^{3}\nabla_{V}X\right)-K\left(V,\,^{3}\nabla_{U}X\right)-K\left(\left[U,V\right],X\right)\label{eq:7}\\
+U\left[K\left(V,X\right)\right]-V\left[K\left(U,X\right)\right] & =\left(\,^{3}\nabla_{U}K\right)\left(V,X\right)-\left(\,^{3}\nabla_{V}K\right)\left(U,X\right)+K\left(\,^{3}T\left(U,V\right),X\right),\nonumber 
\end{align}
with $\,^{3}T\left(U,V\right)$ is the torsion tensor on $\Sigma$.
Inserting (\ref{eq:4}), (\ref{eq:5}), and (\ref{eq:6}) to (\ref{eq:Riemann})
and then using relation (\ref{eq:7}), one obtains the decomposition
of intrinsic curvature on $\mathcal{M}$ into the intrinsic and extrinsic
curvature of $\Sigma$:
\begin{align}
\boldsymbol{R}\left(U,V\right)X= & \,^{3}\boldsymbol{R}\left(U,V\right)X+K\left(V,X\right)\nabla_{U}\hat{n}-K\left(U,X\right)\nabla_{V}\hat{n}\label{eq:8}\\
 & +\left(\left(\,^{3}\nabla_{U}K\right)\left(V,X\right)-\left(\,^{3}\nabla_{V}K\right)\left(U,X\right)+K\left(\,^{3}T\left(U,V\right),X\right)\right)\hat{n}.\nonumber 
\end{align}
$\,^{3}\boldsymbol{R}$ is the generalized Riemann curvature of the
connection $\,^{3}\nabla$ on the hypersurface $\Sigma$. The only
symmetry possessed by the generalized Riemann curvature (\ref{eq:Riemann})
is:
\[
\boldsymbol{R}\left(U,V\right)X=-\boldsymbol{R}\left(V,U\right)X.
\]

As a comment concerning the terminology, the definition in (\ref{eq:Riemann})
is defined originally for the Levi-Civita connection, and so does
the term Riemann curvature tensor. However, in this article, we use
a similar term and definition (i.e., (\ref{eq:Riemann})) for the
intrinsic curvature of a generalized affine connection. From the decomposition
of generalized Riemann tensor (\ref{eq:8}), the Gauss-Codazzi-Mainardi
equations could be obtained.

\subsubsection*{The Acceleration Tensor}

In the Eulerian frame, one could regard the normal $\hat{n}$ as the
4-velocity of an event $p\in\mathcal{M}$ \cite{Eric}. Let us choose
the spatial part $x^{i}$ of $x^{\mu}=\left(x^{0},x^{i}\right)$ as
an 'adapted' local ($n$--1) coordinate on $\varSigma$ (notice that
in general, the spatial part $x^{i}$ of $x^{\mu}=\left(x^{0},x^{i}\right)$
is not necessarily a local coordinate on $\varSigma$). The vector
basis related to this coordinate is $\partial_{i}$. Together with
$\hat{n}$, they define a complete basis on $T_{p}\mathcal{M},$ namely
$\left(\hat{n},\partial_{i}\right)$. Let us define the acceleration
tensor $\boldsymbol{a}$ in the coordinate basis $\left(\hat{n},\partial_{i}\right)$
as follows:
\begin{equation}
\boldsymbol{a}=\boldsymbol{a}_{\nu}=\left(\boldsymbol{a}_{\hat{n}},\boldsymbol{a}_{i}\right):=\left(\nabla_{\hat{n}}\hat{n},\nabla_{i}\hat{n}\right).\label{eq:accell}
\end{equation}
where:
\[
\boldsymbol{a}_{\nu}=\boldsymbol{a}_{\nu}^{\mu}\partial_{\mu}\;\;\in\:T_{p}\mathcal{M},\qquad\nu=\left(\hat{n},i\right),\:i=1,2,..,n-1.
\]
Hence, for an arbitrary vector $U\in T_{p}\Sigma$, one has:
\[
\nabla_{U}\hat{n}=U^{i}\nabla_{i}\hat{n}=U^{i}\boldsymbol{a}_{i}=\boldsymbol{a}\left(U\right).
\]
The quantity $\boldsymbol{a}_{\hat{n}}:=\alpha$ is known as the 4-acceleration.

It is a well-known fact that in the original General Relativity with
Levi-Civita connection, the 4-acceleration is orthogonal to the 4-velocity,
namely $g\left(\alpha,\hat{n}\right)=0$. This is a consequence of
the metric compatibility $\nabla_{X}g=0$, $\forall\:X\in T_{p}\mathcal{M}$.
With the same condition, one could show that the acceleration tensor
is orthogonal to the 4-velocity as well:
\[
g\left(\boldsymbol{a}_{\nu},\hat{n}\right)=g\left(\nabla_{\nu}\hat{n},\hat{n}\right)=0,\qquad\nu=\left(\hat{n},i\right),\:i=1,2,..,n-1,
\]
using the fact that:
\[
\underset{0}{\underbrace{\nabla_{\nu}g\left(\hat{n},\hat{n}\right)}}=\underset{0}{\underbrace{\left(\nabla_{\nu}g\right)}}\left(\hat{n},\hat{n}\right)+g\left(\nabla_{\nu}\hat{n},\hat{n}\right)+g\left(\hat{n},\nabla_{\nu}\hat{n}\right),
\]
and the symmetricity of $g$. Notice that if the metricity condition
is violated, one has:
\[
g\left(\boldsymbol{a}_{\nu},\hat{n}\right)=-\frac{1}{2}\left(\nabla_{\nu}g\right)\left(\hat{n},\hat{n}\right),
\]
which defines the 'angle' between $\boldsymbol{a}_{\nu}$ and $\hat{n}.$
Let us label the angle as:
\begin{equation}
\Theta_{\nu}=\pm g\left(\boldsymbol{a}_{\nu},\hat{n}\right).\label{eq:angle}
\end{equation}
For an arbitrary vector $U\in T_{p}\Sigma$, one could write $\Theta\left(U\right)=\pm g\left(\boldsymbol{a}\left(U\right),\hat{n}\right)$,
that is, the angle between the acceleration in the spatial direction
$U$, with the normal $\hat{n}.$ The definition of the angle $\Theta$
will be useful for the next sections.

\subsubsection*{Extrinsic Curvature of the Second Kind}

As we had mentioned in the previous sections, besides the extrinsic
curvature defined in (\ref{eq:3}), there exists another definition
of the extrinsic curvature \cite{Eric}:
\begin{equation}
\mathcal{K}\left(U,V\right)=g\left(\nabla_{U}\hat{n},V\right),\qquad U,V\in T_{p}\Sigma.\label{eq:extrinsic2}
\end{equation}
For the original General Relativity where the connection is Levi-Civita,
the first and second extrinsic curvature coincide (by the factor of
$\mp1$ for Riemannian/ Lorentzian case):
\[
\underset{0}{\underbrace{\nabla_{U}g\left(V,\hat{n}\right)}}=\underset{0}{\underbrace{\left(\nabla_{U}g\right)}}\left(V,\hat{n}\right)+\underset{\pm K\left(U,V\right)}{\underbrace{g\left(\nabla_{U}V,\hat{n}\right)}}+\underset{\mathcal{K}\left(U,V\right)}{\underbrace{g\left(V,\nabla_{U}\hat{n}\right)}}.
\]
Notice that without the metric compatibility, they differ by $\nabla_{U}g\left(V,\hat{n}\right)$.

Written in terms of components, one could show that the extrinsic
curvature of the second kind satisfies: 
\begin{equation}
\mathcal{K}\left(U,V\right)=g\left(\nabla_{U}\hat{n},V\right)=U^{\mu}V^{\nu}g\left(\nabla_{\mu}\hat{n},\partial_{\nu}\right)=\underset{0}{\underbrace{V_{\alpha}U^{\mu}\partial_{\mu}n^{\alpha}}}+U^{\mu}V_{\alpha}n^{\beta}\omega_{\mu\,\:\,\beta}^{\,\:\,\alpha},\label{eq:ex2index}
\end{equation}
where the first term is zero since $U^{\mu}\partial_{\mu}V_{\alpha}n^{\alpha}$
and $n^{\alpha}U^{\mu}\partial_{\mu}V_{\alpha}$ are zero for $U,V\in T_{p}\Sigma$.
The geometrical interpretation of $\mathcal{K}\left(\partial_{\mu},\partial_{\nu}\right)$
differs from $K\left(\partial_{\mu},\partial_{\nu}\right)$: it describes
the $\nu^{\mathrm{th}}$-component of the change of the normal in
the direction of $\partial_{\mu}$. 

Before we proceed to the main calculation, it is convenient to fix
the notations and conventions we use concerning the metric and the
inner-product. First, the covector $V^{*}\in T_{p}^{*}\mathcal{M}$
is defined as follows:
\begin{equation}
g\left(V,W\right)=\left\langle V^{*},W\right\rangle ,\qquad\forall\;\;W\in T_{p}\mathcal{M},\label{eq:covector}
\end{equation}
with $g=g_{\mu\nu}dx^{\mu}\otimes dx^{\nu}$ is the metric on $\mathcal{M},$
$V\in T_{p}\mathcal{M}$, and $\left\langle \cdot,\cdot\right\rangle $
is an inner product on $T_{p}\mathcal{M}.$ With this definition,
then (\ref{eq:extrinsic2}) could be written as $\mathcal{K}\left(U,V\right)=\left\langle V^{*},\nabla_{U}\hat{n}\right\rangle $,
for example. Second, as we have mentioned at the beginning of this
chapter, one could write (\ref{eq:covector}) as $g\left(V,\cdot\right)=g\left(\cdot,V\right)=V^{*}$.
Moreover, given $V^{*},W^{*}\in T_{p}^{*}\mathcal{M},$ (\ref{eq:covector})
could be written as:
\begin{equation}
g\left(V,W\right)=g^{*}\left(V^{*},W^{*}\right),\label{eq:convention}
\end{equation}
with $g^{*}=g^{\mu\nu}\partial_{\mu}\otimes\partial_{\nu}$ is the
dual of metric $g,$ such that $g\left(g^{*}\right)=\mathbb{I}$ (or
in indices: $g^{\mu\sigma}g_{\sigma\nu}=\delta_{\nu}^{\mu}$). However,
these relations should be used carefully, for example $g\left(\partial_{\mu},\partial_{\nu}\right)=g_{\mu\nu}$,
is not equal to either $\left\langle dx^{\mu},\partial_{\nu}\right\rangle =\delta_{\nu}^{\mu}$
or $g^{*}\left(dx^{\mu},dx^{\nu}\right)=g^{\mu\nu}$. As another example,
even if $g\left(V,\cdot\right)=g\left(\cdot,V\right)=V^{*}$, this
is not valid for basis vectors and covectors, as $g\left(\partial_{\mu}\right)\neq dx^{\mu}$.
The correct relation is:
\begin{equation}
dx^{\mu}=g^{\mu\nu}g\left(\partial_{\nu}\right).\label{eq:pentingx}
\end{equation}

\subsubsection*{The Torsion Decomposition, Gauss, and Codazzi-Mainardi Equations}

The results (\ref{eq:torsion-1}) and (\ref{eq:8}) will be contracted
with vectors $\hat{n}$ and $Y\in T_{p}\Sigma$ to obtain the temporal
and spatial parts of the equations. As a first step, we need to decompose
the quantity $\nabla_{U}\hat{n}$ into the normal and parallel parts.
Let us consider the following quantity:
\begin{equation}
\nabla_{U}\left(g\left(X,\hat{n}\right)\right)=\left(\nabla_{U}g\right)\left(X,\hat{n}\right)+g\left(\nabla_{U}X,\hat{n}\right)+g\left(X,\nabla_{U}\hat{n}\right),\qquad X\in T_{p}\Sigma.\label{eq:b}
\end{equation}
The LHS of (\ref{eq:b}) could be written as $\nabla_{U}\left(g\left(X,\hat{n}\right)\right)=U\left[g\left(X,\hat{n}\right)\right]=0,$
since $X\in T_{p}\Sigma$ is perpendicular to $\hat{n}$. One could
do the following decomposition on the quantity $\nabla_{U}\hat{n}$:
\begin{equation}
\nabla_{U}\hat{n}=\underset{\left(\nabla_{U}\hat{n}\right)_{\perp\Sigma}}{\underbrace{\pm\left\langle \hat{n}^{*},\nabla_{U}\hat{n}\right\rangle \hat{n}}}+\underset{\left(\nabla_{U}\hat{n}\right)_{\parallel\Sigma}}{\underbrace{\left\langle \,^{3}dx^{i},\nabla_{U}\hat{n}\right\rangle \partial_{i}}}.\label{eq:d}
\end{equation}
with $\partial_{i}$ and $\,^{3}dx^{i}=\,^{3}q^{ij}q\left(\partial_{j}\right)$
are the vector and covector basis corresponding to $x^{i},$ the adapted
coordinate on $\varSigma$. With the definitions in (\ref{eq:angle})
and (\ref{eq:extrinsic2}), one could obtain the following relations:
\begin{align*}
\left(\nabla_{U}\hat{n}\right)_{\perp\Sigma} & =\pm\left\langle \hat{n}^{*},\nabla_{U}\hat{n}\right\rangle \hat{n}=\pm g\left(\nabla_{U}\hat{n},\hat{n}\right)\hat{n}=\Theta\left(U\right)\hat{n},\\
\left(\nabla_{U}\hat{n}\right)_{\parallel\Sigma} & =\left\langle \,^{3}dx^{i},\nabla_{U}\hat{n}\right\rangle \partial_{i}=\,^{3}q^{ij}g\left(\partial_{j},\nabla_{U}\hat{n}\right)\partial_{i}=\,^{3}q^{ij}\mathcal{K}\left(U,\partial_{j}\right)\partial_{i},
\end{align*}
using equation (\ref{eq:1a}), (\ref{eq:extrinsic2}), and (\ref{eq:pentingx}).
Inserting (\ref{eq:d}) to (\ref{eq:8}) gives:
\begin{align}
\boldsymbol{R}\left(U,V\right)X & =\,^{3}\boldsymbol{R}\left(U,V\right)X+\,^{3}q^{ij}\left(K\left(V,X\right)\mathcal{K}\left(U,\partial_{j}\right)-K\left(U,X\right)\mathcal{K}\left(V,\partial_{j}\right)\right)\partial_{i}\label{eq:e}\\
 & +\left(\left(\,^{3}\nabla_{U}K\right)\left(V,X\right)-\left(\,^{3}\nabla_{V}K\right)\left(U,X\right)+K\left(\,^{3}T\left(U,V\right),X\right)+K\left(\Theta\left(U\right)V-\Theta\left(V\right)U,X\right)\right)\hat{n}.\nonumber 
\end{align}

For the next step, by contracting (\ref{eq:e}) with $\hat{n}$, we
obtain the generalized Codazzi-Mainardi equation:
\begin{align}
g\left(\hat{n},\boldsymbol{R}\left(U,V\right)X\right)= & \pm\left(\left(\,^{3}\nabla_{U}K\right)\left(V,X\right)-\left(\,^{3}\nabla_{V}K\right)\left(U,X\right)+K\left(\,^{3}T\left(U,V\right),X\right)+K\left(\Theta\left(U\right)V-\Theta\left(V\right)U,X\right)\right).\label{eq:CM}
\end{align}
Moreover, contracting (\ref{eq:e}) by $Y\in T_{p}\Sigma$ gives the
generalized Gauss equation:
\begin{eqnarray}
g\left(Y,\boldsymbol{R}\left(U,V\right)X\right) & = & g\left(Y,\,^{3}\boldsymbol{R}\left(U,V\right)X\right)+K\left(V,X\right)\mathcal{K}\left(U,Y\right)-K\left(U,X\right)\mathcal{K}\left(V,Y\right),\label{eq:i}
\end{eqnarray}
where $Y$ could be written in the adapted coordinate as $Y=Y^{i}\partial_{i}$,
with $g\left(Y^{j}\partial_{j},\partial_{i}\right)=\,^{3}Y_{i}$,
and $Y^{*}=\,^{3}Y_{i}\,^{3}dx^{i}\in T_{p}^{*}\Sigma$. Equation
(\ref{eq:torsion-1}) could be equally contracted with $\hat{n}$
and $Z\in T_{p}\Sigma$ to give the torsion decomposition relations:
\begin{eqnarray}
g\left(\hat{n},T\left(X,Y\right)\right) & = & \left(K\left(X,Y\right)-K\left(Y,X\right)\right),\label{eq:torsion1}\\
g\left(Z,T\left(X,Y\right)\right) & = & g\left(Z,\,^{3}T\left(X,Y\right)\right).\label{eq:torsion2}
\end{eqnarray}
These four equations are the main results in this article and will
be heavily used in the next sections. Notice that if the connection
is metric and torsionless, then $K\left(U,V\right)=\mp\mathcal{K}\left(U,V\right)$,
$\Theta\left(U\right)=0$, and $T\left(U,V\right)=0,$ hence the generalized
GCM returns to the standard original GCM.

\section{Metric, Torsionless, and Levi-Civita Cases}

In this section, we will discuss some special cases of the affine
connection, in particular, metric, torsionless, and Levi-Civita connection.
We will show that the conditions applied on the connection $\nabla$
on $\mathcal{M}$ are inherited to the connection $\,^{3}\nabla$
on the hypersurface $\Sigma$.

To prevent any confusion, we need to clarify the conventions we use
to write the notations indices of some geometrical objects. This is
important because of the different conventions used in the literature.
First, we write, using indices, the (vector potential of the) connection
as $\nabla_{\mu}\partial_{\beta}=\omega_{\mu\,\:\,\beta}^{\,\:\,\alpha}\partial_{\alpha}$,
the torsion tensor as $\boldsymbol{T}\left(\partial_{\mu},\partial_{\beta}\right)=T_{\mu\,\:\,\beta}^{\:\:\alpha}\partial_{\alpha}$,
and the generalized Riemann tensor as $\boldsymbol{R}\left(\partial_{\mu},\partial_{\nu}\right)\partial_{\beta}=R_{\mu\nu\,\:\,\,\beta}^{\:\:\:\:\,\,\alpha}\partial_{\alpha}.$
These conventions are different from the ones used in, for examples
\cite{Sotiriou,Olmo}.

\subsubsection*{Metric Compatibility}

A metric connection is a connection that satisfies the metric compatibility
condition:
\begin{equation}
\nabla_{X}g=X^{\mu}\nabla_{\mu}\left(g_{\alpha\beta}dx^{\alpha}dx^{\beta}\right)=0,\qquad\forall\,X\,\in T_{p}\mathcal{M}.\label{eq:metric}
\end{equation}
The condition could be written in components, with abuse of notation
as follows:
\begin{equation}
\nabla_{\mu}g_{\alpha\beta}=\partial_{\mu}g_{\alpha\beta}-\omega_{\mu\beta\alpha}-\omega_{\mu\alpha\beta}=0.\label{eq:z00}
\end{equation}
It describes the failure of the last two indices of $\omega$ (borrowing
the term from \cite{Miller}, the 'rotation' bivector indices) to
be antisymmetric, differing with the quantity $\partial_{\mu}g_{\alpha\beta}$.
In the next paragraph, we will show that the metric compatibility
of a connection in $\mathcal{M}$ will induce also a metric connection
in $\Sigma$.

Let us remember that the metric $g$ in $\mathcal{M}$ could be decomposed
into the perpendicular and parallel parts with respect to the embedded
hypersurface $\Sigma,$ as in (\ref{eq:1a}):
\begin{equation}
g=\underset{\,^{3}q}{\underbrace{g_{\parallel}}}\pm\underset{\hat{n}^{*}\otimes\hat{n}^{*}}{\underbrace{g_{\perp}}}.\label{eq:z-0}
\end{equation}
With this, we could obtain the following relation:
\begin{equation}
\nabla_{X}g=\underset{\nabla_{X}\,^{3}q}{\underbrace{\nabla_{X}g_{\parallel}}}\pm\nabla_{X}g_{\perp}=0,\qquad X\in T_{p}\Sigma,\label{eq:z-1}
\end{equation}
from the metric compatibility. On the other hand, we could directly
decompose $\nabla_{X}g$ as:
\[
\nabla_{X}g=\left(\nabla_{X}g\right)_{\parallel}\pm\left(\nabla_{X}g\right)_{\perp}=0,\quad\implies\quad\left(\nabla_{X}g\right)_{\parallel}=0,\quad\left(\nabla_{X}g\right)_{\perp}=0.
\]
Notice that in general, $\left(\nabla_{X}g\right)_{\parallel}\neq\nabla_{X}g_{\parallel}$
and $\left(\nabla_{X}g\right)_{\perp}\neq\nabla_{X}g_{\perp}$. Now
let us consider the following quantity:
\begin{equation}
\nabla_{X}\,^{3}q=\left(\nabla_{X}\,^{3}q\right)_{\parallel}\pm\left(\nabla_{X}\,^{3}q\right)_{\perp}.\label{eq:z-2}
\end{equation}
Since $X\in T_{p}\varSigma$ and $\left(\nabla_{X}\,^{3}q\right)_{\parallel}$
live in the hypersurface $\Sigma$, we could define:
\begin{equation}
\left(\nabla_{X}\,^{3}q\right)_{\parallel}:=\,^{3}\nabla_{X}\,^{3}q,\label{eq:z-3a}
\end{equation}
with $\,^{3}\nabla$ is the induced connection on hypersurface $\Sigma$.
With the definition (\ref{eq:z-3a}) and (\ref{eq:z-1}), we obtain:
\[
\,^{3}\nabla_{X}\,^{3}q=\mp\nabla_{X}g_{\perp}\pm\left(\nabla_{X}g_{\perp}\right)_{\perp}.
\]
Decomposing $\nabla_{X}g_{\perp}$ in its parallel and perpendicular
part (and also using (\ref{eq:z-1})), we have:
\[
\,^{3}\nabla_{X}\,^{3}q=\mp\left(\nabla_{X}g_{\perp}\right)_{\parallel}.
\]
Using (\ref{eq:z-0}), then one obtains:
\begin{equation}
\,^{3}\nabla_{X}\,^{3}q=\mp\left(\nabla_{X}\hat{n}^{*}\otimes\hat{n}^{*}\right)_{\parallel}=\mp\left(\left(\nabla_{X}\hat{n}^{*}\right)\otimes\hat{n}^{*}+\hat{n}^{*}\otimes\nabla_{X}\hat{n}^{*}\right)_{\parallel}.\label{eq:135}
\end{equation}
Notice that the RHS of (\ref{eq:135}) could be written as a projection
of such quantity to $\Sigma$ using a projection operator $p_{\Sigma}=g^{\mu\lambda}\,^{3}q_{\lambda\nu}\partial_{\mu}\otimes dx^{\nu}$,
namely:
\[
\left(\left(\nabla_{X}\hat{n}^{*}\right)\otimes\hat{n}^{*}+\hat{n}^{*}\otimes\nabla_{X}\hat{n}^{*}\right)_{\parallel}=p_{\Sigma}\cdot\left(\left(\nabla_{X}\hat{n}^{*}\right)\otimes\hat{n}^{*}+\hat{n}^{*}\otimes\nabla_{X}\hat{n}^{*}\right)\cdot p_{\Sigma}^{T},
\]
with $\cdot$ denotes the inner product (matrix multiplication) between
matrices. One could write:
\begin{align*}
p_{\Sigma}\cdot\left(\hat{n}^{*}\otimes\nabla_{X}\hat{n}^{*}\right) & =p_{\Sigma}\left(\hat{n}^{*}\right)\otimes\nabla_{X}\hat{n}^{*},\\
\left(\left(\nabla_{X}\hat{n}^{*}\right)\otimes\hat{n}^{*}\right)\cdot p_{\Sigma}^{T} & =\left(\nabla_{X}\hat{n}^{*}\right)\otimes p_{\Sigma}\left(\hat{n}^{*}\right).
\end{align*}
Since $p_{\Sigma}\left(\hat{n}^{*}\right)=0$ ($\,^{3}q$ lives in
$\Sigma$ and $\hat{n}$ is perpendicular to $\Sigma$ ), we have:
\[
\,^{3}\nabla_{X}\,^{3}q=0.
\]
Therefore, we have shown that the metric compatibility of a connection
$\nabla$ on $\left(\mathcal{M},g\right),$ induces a connection $\,^{3}\nabla$
on the hypersurface $\left(\Sigma,\,^{3}q\right),$ which also satisfies
the metric compatibility with respect to $\,^{3}q$.

\subsubsection*{Torsionless Condition}

Let us perform a similar derivation for a special condition on the
torsion tensor. The torsionless condition is a case where the connection
has zero torsion:
\begin{equation}
T\left(X,Y\right)=0,\qquad\forall\,X,Y\,\in T_{p}\mathcal{M}.\label{eq:torsionles}
\end{equation}
With the definition of torsion in (\ref{eq:torsion}), a direct consequence
is:
\begin{equation}
T\left(\partial_{\alpha},\partial_{\beta}\right)=\underset{T_{\alpha\:\:\beta}^{\,\:\,\sigma}}{\underbrace{\left(\omega_{\alpha\,\:\beta}^{\,\:\:\sigma}-\omega_{\beta\,\:\alpha}^{\:\:\:\sigma}\right)}}\partial_{\sigma}=0,\qquad\omega_{\alpha\,\:\beta}^{\,\:\:\sigma}=\omega_{\beta\,\:\alpha}^{\:\:\:\sigma},\label{eq:z-3}
\end{equation}
which results in the symmetricity of the first and third index of
the connection. Notice here that we do not use factor $2$ on the
definition of torsion, as in \cite{Sotiriou,Olmo}. Using the torsion
decomposition (\ref{eq:torsion-1}), the torsionless condition becomes:
\[
\,^{3}T\left(X,Y\right)+\left(K\left(X,Y\right)-K\left(Y,X\right)\right)\hat{n}=0.
\]
From the definition of the first extrinsic curvature (\ref{eq:z}),
and then using (\ref{eq:z-3}), we have:
\[
\,^{3}T\left(X,Y\right)=0.
\]
Hence, the torsionless connection $\nabla$ on $\mathcal{M}$ induced
a torsionless connection $\,^{3}\nabla$ on $\Sigma$.

\subsubsection*{Levi-Civita Connection}

Since the metric compatibility and torsionless condition are independent
of one another, one could define a connection that satisfies both
of the conditions. This connection is unique and known as the Levi-Civita
connection. A metric compatible and torsionless connection $\nabla$
on $\left(\mathcal{M},g\right)$ induces a connection $\,^{3}\nabla$
on $\left(\Sigma,\,^{3}q\right)$ which is also metric (with respect
to $\,^{3}q$) and torsionless.

The Levi-Civita connection could be derived as follows. With the metric
compatibility condition (\ref{eq:z00}), one considers a certain combination
of the vector potential satisfies equality as follows:
\begin{equation}
\omega_{\mu\alpha\beta}+\omega_{\mu\beta\alpha}-\omega_{\alpha\beta\mu}-\omega_{\alpha\mu\beta}+\omega_{\beta\mu\alpha}+\omega_{\beta\alpha\mu}=\partial_{\mu}g_{\alpha\beta}-\partial_{\alpha}g_{\beta\mu}+\partial_{\beta}g_{\mu\alpha}.\label{eq:quantity}
\end{equation}
Using the torsionless condition (\ref{eq:z-3}) on the LHS, one could
obtain, and then define the (vector potential of the) Levi-Civita
connection:
\begin{equation}
\omega_{\mu\alpha\beta}:=\Gamma_{\mu\alpha\beta}=\frac{1}{2}\left(\partial_{\mu}g_{\alpha\beta}-\partial_{\alpha}g_{\beta\mu}+\partial_{\beta}g_{\mu\alpha}\right),\label{eq:LevCiv}
\end{equation}
also known as the Christoffel symbol. It is easy to show that the
Levi-Civita connection is symmetric on the first and third indices,
namely $\Gamma_{\mu\alpha\beta}=\Gamma_{\beta\alpha\mu}.$

\subsubsection*{The Symmetric and Antisymmetric Parts of Connections, Torsion, and
Curvatures}

A general affine connection could be decomposed into its Levi-Civita,
torsion, and non-metricity part. The derivation could be found in
classic literature, for example, in \cite{Schouten}. Let us consider
the previous combinations of $\omega$ as in (\ref{eq:quantity}),
but now without the metric compatibility. With (\ref{eq:metric}),
one has: 
\[
\omega_{\mu\alpha\beta}+\omega_{\mu\beta\alpha}-\omega_{\alpha\beta\mu}-\omega_{\alpha\mu\beta}+\omega_{\beta\mu\alpha}+\omega_{\beta\alpha\mu}=\partial_{\mu}g_{\alpha\beta}-\nabla_{\mu}g_{\alpha\beta}-\partial_{\alpha}g_{\beta\mu}+\nabla_{\alpha}g_{\beta\mu}+\partial_{\beta}g_{\mu\alpha}-\nabla_{\beta}g_{\mu\alpha}.
\]
Notice here that we do not use any requirement for the equation. Using
(\ref{eq:torsion}) to rewrite the LHS, one obtains:
\[
2\omega_{\mu\alpha\beta}+T_{\mu\beta\alpha}+T_{\beta\mu\alpha}+T_{\beta\alpha\mu}=\partial_{\mu}g_{\alpha\beta}-\partial_{\alpha}g_{\beta\mu}+\partial_{\beta}g_{\mu\alpha}-\left(\nabla_{\mu}g_{\alpha\beta}-\nabla_{\alpha}g_{\beta\mu}+\nabla_{\beta}g_{\mu\alpha}\right),
\]
that could be written as:
\begin{equation}
\omega_{\mu\alpha\beta}=\Gamma_{\mu\alpha\beta}+C_{\mu\alpha\beta}+Q_{\mu\alpha\beta},\label{eq:Decomposition}
\end{equation}
with the Levi-Civita connection $\Gamma$ satisfies (\ref{eq:LevCiv}),
and:
\begin{align}
Q_{\mu\alpha\beta} & =-\frac{1}{2}\left(\nabla_{\mu}g_{\alpha\beta}-\nabla_{\alpha}g_{\beta\mu}+\nabla_{\beta}g_{\mu\alpha}\right),\label{eq:metricpart}\\
C_{\mu\alpha\beta} & =\frac{1}{2}\left(T_{\mu\alpha\beta}+T_{\alpha\beta\mu}-T_{\beta\mu\alpha}\right).\label{eq:connectionpart}
\end{align}
$Q$ and $C$ are the non-metricity and the torsion parts, usually
known respectively, as the disformation/deflection and the contorsion
tensor. If the disformation $Q$ is zero, the connection is metric,
and if the contorsion $C$ is zero, the connection is torsionless.
If both of them are zero, the connection is Levi-Civita.

Let $\left(\alpha\right|...\left|\beta\right)$ and $\left[\alpha\right|...\left|\beta\right]$
denote, respectively, the symmetric and antisymmetric part of an arbitrary
tensor with indices $\alpha$ and $\beta$. With these notations,
one could write the symmetries of $\Gamma$, $Q$ and $C$ as follows:
\begin{center}
\begin{tabular}{lclcl}
$\Gamma_{\left(\mu\alpha\right)\beta}=\frac{1}{2}\partial_{\beta}g_{\mu\alpha},$ & $\qquad$ & $Q_{\left(\mu\alpha\right)\beta}=-\frac{1}{2}\nabla_{\beta}g_{\mu\alpha},$ & $\qquad$ & $C_{\left(\mu\alpha\right)\beta}=\frac{1}{2}\left(T_{\mu\alpha\beta}-T_{\beta\mu\alpha}\right)$\tabularnewline
$\Gamma_{\left[\mu\alpha\right]\beta}=\partial_{\left[\mu\right.}g_{\left.\alpha\right]\beta},$ &  & $Q_{\left[\mu\alpha\right]\beta}=-\nabla_{\left[\mu\right.}g_{\left.\alpha\right]\beta},$ &  & $C_{\left[\mu\alpha\right]\beta}=\frac{1}{2}T_{\alpha\beta\mu},$\tabularnewline
$\Gamma_{\left(\mu\right|\alpha\left|\beta\right)}=\Gamma_{\mu\alpha\beta},$ &  & $Q_{\left(\mu\right|\alpha\left|\beta\right)}=Q_{\mu\alpha\beta},$ &  & $C_{\left(\mu\right|\alpha\left|\beta\right)}=\frac{1}{2}\left(T_{\alpha\beta\mu}-T_{\beta\mu\alpha}\right),$\tabularnewline
$\Gamma_{\left[\mu\right|\alpha\left|\beta\right]}=0,$ &  & $Q_{\left[\mu\right|\alpha\left|\beta\right]}=0,$ &  & $C_{\left[\mu\right|\alpha\left|\beta\right]}=\frac{1}{2}T_{\mu\alpha\beta},$\tabularnewline
$\Gamma_{\mu\left(\alpha\beta\right)}=\frac{1}{2}\partial_{\mu}g_{\alpha\beta},$ &  & $Q_{\mu\left(\alpha\beta\right)}=-\frac{1}{2}\nabla_{\mu}g_{\alpha\beta},$ &  & $C_{\mu\left(\alpha\beta\right)}=0,$\tabularnewline
$\Gamma_{\mu\left[\alpha\beta\right]}=-\partial_{\left[\alpha\right.}g_{\left.\beta\right]\mu},$ &  & $Q_{\mu\left[\alpha\beta\right]}=\nabla_{\left[\alpha\right.}g_{\left.\beta\right]\mu},$ &  & $C_{\mu\left[\alpha\beta\right]}=C_{\mu\alpha\beta}.$\tabularnewline
\end{tabular}
\par\end{center}

From these relations, it is clear that $\Gamma$ and $Q$ have similar
symmetries, in particular, they are symmetric in the first and third
indices. On the other hand, the contorsion $C$ is antisymmetric in
the second and third indices. One should keep in mind that other authors
use different conventions for the order of the indices, in particular
\cite{Sotiriou,Olmo}, but these do not alter the physical symmetries
of the connection.

With the decomposition of connection (\ref{eq:Decomposition}), one
could obtain directly the decomposition of extrinsic curvature of
the first kind:
\[
K\left(X,Y\right)=\underset{K_{\Gamma}\left(X,Y\right)}{\underbrace{\pm X^{\mu}Y^{\beta}\Gamma_{\mu\,\:\beta}^{\:\:\alpha}n_{\alpha}}}+\underset{K_{Q}\left(X,Y\right)}{\underbrace{\pm X^{\mu}Y^{\beta}Q_{\mu\,\:\,\beta}^{\:\:\,\alpha}n_{\alpha}}}+\underset{K_{C}\left(X,Y\right)}{\underbrace{\pm X^{\mu}Y^{\beta}C_{\mu\,\:\,\beta}^{\:\:\,\alpha}n_{\alpha}}},
\]
where $K_{\Gamma}$ and $K_{Q}$ are symmetric on the first and second
argument, hence, the torsionless condition $C=0$, gives a symmetric
extrinsic curvature. On the other hand, the second extrinsic curvature
satisfies the following decomposition:
\[
\mathcal{K}\left(X,Y\right)=\underset{\mathcal{K}_{\Gamma}\left(X,Y\right)}{\underbrace{X^{\mu}Y_{\alpha}n^{\beta}\Gamma_{\mu\,\:\,\beta}^{\,\:\,\alpha}}}+\underset{\mathcal{K}_{Q}\left(X,Y\right)}{\underbrace{X^{\mu}Y_{\alpha}n^{\beta}Q_{\mu\,\:\,\beta}^{\,\:\,\alpha}}}+\underset{\mathcal{K}_{C}\left(X,Y\right)}{\underbrace{X^{\mu}Y_{\alpha}n^{\beta}C_{\mu\,\:\,\beta}^{\,\:\,\alpha}}}.
\]
Notice that now $\mathcal{K}_{\Gamma}$ and $\mathcal{K}_{Q}$ are
\textit{not} symmetric on the first and second argument. However,
for a metric connection, the extrinsic curvatures could be written
as:
\[
K\left(X,Y\right)=\mp\left(\underset{0}{\underbrace{\left(\nabla_{X}g\right)}}\left(Y,n\right)+g\left(Y,\nabla_{X}n\right)\right)=\mp\mathcal{K}\left(X,Y\right),
\]
which comes from relation (\ref{eq:b}). Hence, the extrinsic curvature
of the first kind coincides with the second kind.

The torsion could be written as:
\begin{equation}
T\left(X,Y\right)=\underset{0}{\underbrace{T_{\Gamma}\left(X,Y\right)}}+T_{C}\left(X,Y\right)+\underset{0}{\underbrace{T_{Q}\left(X,Y\right)}}=X^{\mu}Y^{\beta}\left(C_{\mu\,\:\beta}^{\:\:\alpha}-C_{\beta\,\:\mu}^{\:\:\alpha}\right)\partial_{\alpha},\label{eq:Decomposition3}
\end{equation}
from the symmetries of the parts of the affine connection.

The Riemann curvature (\ref{eq:Riemann}) could be rewritten in indices
as follows:
\begin{align}
\boldsymbol{R}\left(\partial_{\mu},\partial_{\mu}\right)\partial_{\beta}= & \nabla_{\mu}\nabla_{\nu}\partial_{\beta}-\nabla_{\nu}\nabla_{\mu}\partial_{\beta}=\underset{R_{\mu\nu\,\:\,\beta}^{\:\:\:\:\,\,\alpha}}{\underbrace{\left(\partial_{\mu}\omega_{\nu\,\:\beta}^{\:\:\alpha}-\partial_{\nu}\omega_{\mu\,\:\beta}^{\:\:\alpha}+\omega_{\mu\,\:\sigma}^{\:\:\alpha}\omega_{\nu\,\:\beta}^{\:\:\sigma}-\omega_{\nu\,\:\sigma}^{\:\:\alpha}\omega_{\mu\,\:\beta}^{\:\:\sigma}\right)}\partial_{\alpha}.}\label{eq:Rindices}
\end{align}
One could insert (\ref{eq:Decomposition}) to (\ref{eq:Rindices}),
to obtain the decomposition of the Riemann tensor in terms of $\Gamma$,
$Q$, and $C$, but it is not necessary for the derivation in this
article. Unlike the first extrinsic curvature and the torsion tensor,
the Riemann tensor could not be cleanly decomposed because of the
existence of the non-linear terms in (\ref{eq:Rindices}).

It is convenient to lower all the indices of the Riemann tensor to
obtain clearer symmetries:
\begin{equation}
R_{\mu\nu\alpha\beta}=\partial_{\mu}\omega_{\nu\alpha\beta}-\partial_{\nu}\omega_{\mu\alpha\beta}+\omega_{\mu\,\:\beta}^{\,\:\:\lambda}\omega_{\nu\lambda\alpha}-\omega_{\nu\,\:\beta}^{\,\:\:\lambda}\omega_{\mu\lambda\alpha}+\omega_{\mu\,\:\beta}^{\,\:\:\lambda}\nabla_{\nu}g_{\lambda\alpha}-\omega_{\nu\,\:\beta}^{\,\:\:\lambda}\nabla_{\mu}g_{\lambda\alpha}.\label{eq:b-2}
\end{equation}
Notice that the generalized Riemann tensor (\ref{eq:Rindices}) only
satisfies the antisymmetricity on the first and second index, namely,
on the loop orientation indices \cite{Miller}: $R_{\mu\nu\alpha\beta}=-R_{\nu\mu\alpha\beta}.$
Only if the connection is metric; $Q=0$, one could show that the
curvature satisfies an additional symmetry, namely $R_{\mu\nu\alpha\beta}=-R_{\mu\nu\beta\alpha}.$
Moreover, if the metric connection is also torsionless, it is symmetric
by the switching between the loop orientation indices $\left(\mu,\nu\right)$
and the 'rotation' bivector indices $\left(\alpha,\beta\right)$,
thus recovering the usual symmetries of the Riemannian tensor of the
Levi-Civita connection: $R_{\mu\nu\alpha\beta}=R_{\alpha\beta\mu\nu},$
where the loop orientation and the rotation bivector indices are symmetric.

\subsubsection*{Bianchi Identities}

The generalized Riemann tensor satisfies the second Bianchi Identity:
For any vector $X,Y,Z\in T_{p}\mathcal{M},$ the second Bianchi Identity
is:
\begin{equation}
\left(\nabla_{X}\boldsymbol{R}\right)\left(Y,Z\right)+\left(\nabla_{Y}\boldsymbol{R}\right)\left(Z,X\right)+\left(\nabla_{Z}\boldsymbol{R}\right)\left(X,Y\right)=0.\label{eq:Bianchi2}
\end{equation}
It is a consequence of the orthogonality of $\omega$ and $\boldsymbol{R}$,
namely, $\mathrm{d}_{\nabla}\boldsymbol{R}=\mathrm{d}_{\nabla}^{2}\omega=\boldsymbol{R}\wedge\omega=0$,
with $\mathrm{d}_{\nabla}$ is the exterior covariant derivative of
the connection $\nabla$, see \cite{Baez}. The Bianchi Identity (\ref{eq:Bianchi2})
could be written in terms of torsion as follows:
\begin{align*}
\left(\left(\nabla_{X}\boldsymbol{R}\right)\left(Y,Z\right)+\left(\nabla_{Y}\boldsymbol{R}\right)\left(Z,X\right)+\left(\nabla_{Z}\boldsymbol{R}\right)\left(X,Y\right)\right)= & \quad\nabla_{\left[X,\left[Y,Z\right]\right]+\left[Y,\left[Z,X\right]\right]+\left[Z,\left[X,Y\right]\right]}W\\
 & +\left(\boldsymbol{R}\left(X,T\left(Y,Z\right)\right)+\boldsymbol{R}\left(Y,T\left(Z,X\right)\right)+\boldsymbol{R}\left(Z,T\left(X,Y\right)\right)\right)W.
\end{align*}
The first term contains the Jacobi identity, which is zero: $\left[X,\left[Y,Z\right]\right]+\left[Y,\left[Z,X\right]\right]+\left[Z,\left[X,Y\right]\right]=0.$
Hence, we have the second Bianchi Identity in terms of torsion:
\[
\boldsymbol{R}\left(X,T\left(Y,Z\right)\right)+\boldsymbol{R}\left(Y,T\left(Z,X\right)\right)+\boldsymbol{R}\left(Z,T\left(X,Y\right)\right)=0.
\]

A torsionless connection satisfies another similar identity, the first
Bianchi Identity. One could show that the following relation is true:
\begin{align*}
R\left(X,Y\right)Z+R\left(Y,Z\right)X+R\left(Z,X\right)Y= & \nabla_{X}T\left(Y,Z\right)+\nabla_{Y}T\left(Z,X\right)+\nabla_{Z}T\left(X,Y\right)+T\left(X,\left[Y,Z\right]\right)\\
 & +T\left(Y,\left[Z,X\right]\right)+T\left(Z,\left[X,Y\right]\right)-\left[X,\left[Y,Z\right]\right]-\left[Y,\left[Z,X\right]\right]-\left[Z,\left[X,Y\right]\right].
\end{align*}
With the Jacobi Identity and the torsionless condition, one obtain
the first Bianchi Identity:
\[
R\left(X,Y\right)Z+R\left(Y,Z\right)X+R\left(Z,X\right)Y=0.
\]

\subsubsection*{Contractions of the Generalized Riemann Tensor}

To understand the significance of Gauss-Codazzi equations for General
Relativity, it is convenient to introduce the contractions of Riemann
tensors, the Ricci tensor $\mathbf{Ric}$ and Ricci scalar $\mathcal{R}$
as follows:
\begin{align}
\mathbf{Ric} & =\left\langle dx^{\mu},\boldsymbol{R}\left(\partial_{\mu},\partial_{\nu}\right)\partial_{\beta}\right\rangle dx^{\nu}\otimes dx^{\beta}=\underset{R_{\nu\beta}}{\underbrace{\delta_{\alpha}^{\mu}R_{\mu\nu\,\:\,\beta}^{\:\:\:\:\,\,\alpha}}}dx^{\nu}\otimes dx^{\beta},\label{eq:ricci}\\
\mathcal{R} & =\mathrm{tr}\,\mathbf{Ric}=g^{\nu\beta}R_{\nu\beta}.\label{eq:scalar}
\end{align}
The lack of symmetries of the generalized Riemann tensor results in
the ambiguities in defining the Ricci tensor. Contracting the index
$\nu$ with $\alpha$ of the Riemann tensor $R_{\mu\nu\,\:\,\beta}^{\:\:\:\:\:\alpha}$
will give the minus of (\ref{eq:ricci}), but contracting either $\mu,$
or $\nu$ with $\beta$ instead of $\alpha$ will give a different
quantity due to the lack of symmetry in the indices $\left(\alpha,\beta\right)$.
This quantity is defined as the co-Ricci tensor \cite{Jimenez2}:
\begin{equation}
\mathbf{\overline{Ric}}=\underset{R_{\nu\alpha}}{\underbrace{g^{\mu\beta}R_{\mu\nu\alpha\beta}}}dx^{\nu}\otimes dx^{\alpha}.\label{eq:coRic}
\end{equation}
Another type of contraction also exists, known as the homothetic curvature
\cite{Iosifidis3,Jimenez2,Iosifidis4}:
\[
\mathrm{tr}\boldsymbol{R}\left(\partial_{\mu},\partial_{\nu}\right)=\delta_{\alpha}^{\beta}R_{\mu\nu\,\:\,\beta}^{\:\:\:\:\:\alpha},
\]
this, again, is not zero as in the Levi-Civita case, due to the non-metricity
nature of the general affine connection.

Moreover, the Ricci tensor is no longer symmetric in the indices $\left(\nu,\beta\right)$.
This will cause problem in constructing the Einstein tensor $\boldsymbol{G}$,
that needs to be symmetric since it is proportional to the stress-energy
tensor $\mathcal{T}$ via the Einstein-Field Equation $\boldsymbol{G}=\kappa\mathcal{T}.$
However, in Metric-Affine and $f\left(\mathcal{R}\right)$-gravity,
only the symmetric part of the Ricci tensor enters the equation of
motion \cite{Sotiriou}.

\section{Application to Gravity}

\subsection{The Modified Theories of Gravity}

\subsubsection*{Second Order Formulation: The Standard GR and $f\left(\mathcal{R}\right)$-Gravity}

The original derivation of Einstein Field Equation (EFE) from a variational
principle was first considered by Einstein and Hilbert, where the
action is $S_{EH}\left[g\right]=\mathcal{R}\textrm{vol}+S_{\mathrm{matter}}\left[g\right]$,
as a functional of the metric $g.$ The connection is assumed \textit{a
priori} to be Levi-Civita and therefore is a function of the metric.
Minimizing the action with respect to the variation of $g$ gives
the standard EFE. This is known as the second-order formulation of
gravity.

$f\left(\mathcal{R}\right)$-gravity is one of the modification of
General Relativity via a more general choice of action:
\begin{equation}
S\left[g\right]=\intop_{\mathcal{M}}f\left(\mathcal{R}\right)\mathrm{vol}+S_{\mathrm{matter}}\left[g\right],\label{eq:fR}
\end{equation}
where $f\left(\mathcal{R}\right)$ is the power series function of
Ricci scalar with the form as follows \cite{Oikonomou,Sotiriou}:
\[
f\left(\mathcal{R}\right)=..+\frac{c_{2}}{\mathcal{R}^{2}}+\frac{c_{1}}{\mathcal{R}}-2\Lambda+\mathcal{R}+\frac{\mathcal{R}^{2}}{k_{2}}+\frac{\mathcal{R}^{3}}{k_{3}}+...
\]
Minimizing the action (\ref{eq:fR}) with respect to the metric gives
the $f\left(\mathcal{R}\right)$-EFE as follows:
\begin{equation}
f'\left(\mathcal{R}\right)R_{\left(\alpha\beta\right)}-\frac{1}{2}f\left(\mathcal{R}\right)g_{\alpha\beta}=\kappa\mathcal{T}_{\alpha\beta}.\label{eq:EL1}
\end{equation}
$\mathcal{T}_{\alpha\beta}$ is the stress-energy-momentum tensor,
obtained from the variation of $g$ on $S_{\mathrm{matter}}$:
\begin{equation}
\delta_{g}S_{\mathrm{matter}}\left[g\right]=-\intop_{\mathcal{M}}\kappa\mathcal{T}_{\alpha\beta}\delta g^{\alpha\beta}\mathrm{vol}.\label{eq:tiga}
\end{equation}
Standard GR is a special case of $f\left(\mathcal{R}\right)$-gravity,
where $f\left(\mathcal{R}\right)=\mathcal{R}$.

\subsubsection*{First Order Formulation: The Palatini Formalism}

In a more general approach known as the Palatini formalism (or the
first-order formulation of gravity), the connection is not assumed
to be Levi-Civita, hence it is independent of the metric on $\mathcal{M}$
\cite{Olmo}. The Einstein-Hilbert action is now a functional over
the metric $g$ and connection $\omega$:
\begin{equation}
S_{EH}\left[g,\omega\right]=\intop_{\mathcal{M}}\mathcal{R}\mathrm{vol}+S_{\mathrm{matter}}\left[g\right].\label{eq:Palatini}
\end{equation}
$\omega$ is a general affine connection defined in (\ref{eq:Decomposition}).
Minimizing the action with respect to the variation of $g$ gives
an equation of motion similar to the EFE:
\begin{equation}
\underset{G_{\alpha\beta}}{\underbrace{R_{\left(\alpha\beta\right)}-\frac{1}{2}\mathcal{R}g_{\alpha\beta}}}=\kappa\mathcal{T}_{\alpha\beta},\label{eq:EL1-1}
\end{equation}
but with the torsional and non-metricity contributions inside $R_{\left[\alpha\beta\right]}$
and $\mathcal{R}$. Minimizing the action (\ref{eq:Palatini}) with
respect to the connection $\omega$ results in another equation of
motion:
\begin{equation}
\underset{P_{\mu}^{\:\:\alpha\beta}}{\underbrace{\left(\underset{A_{\nu}}{\underbrace{T_{\lambda\,\:\nu}^{\:\:\lambda}-\frac{1}{2}g^{\sigma\lambda}\nabla_{\nu}g_{\sigma\lambda}}}-\nabla_{\nu}\right)\left(2\delta_{\mu}^{\:\left[\nu\right.}g^{\left.\alpha\right]\beta}\right)+T_{\mu}^{\:\:\alpha\beta}}}=0.\label{eq:EL2}
\end{equation}
$G_{\alpha\beta}$ and $P_{\mu}^{\:\:\alpha\beta}$ are known, respectively,
as the Einstein and Palatini tensor \cite{Iosifidis4}. Although (\ref{eq:EL2})
is different from the result found in the literature \cite{Sotiriou,Iosifidis4,Olmo},
it could be shown that they are equal, up to the factor 2 in the third
term. This is due to the difference in the definition of torsion in
(\ref{eq:torsion}) which does not include the factor $\frac{1}{2},$
as defined in literatures, in particular \cite{Sotiriou,Iosifidis4,Olmo}.

As clearly derived in \cite{Iosifidis4}, the solution to (\ref{eq:EL2})
contains unspecified vector degrees of freedom on the connection.
However, since the Einstein-Hilbert action (\ref{eq:Palatini}) is
invariant under a projective transformation, these vector degrees
of freedom could be eliminated, which results in a condition that
the connection must be metric compatible and torsionless, hence Levi-Civita
\cite{Sotiriou,Olmo}. This will be discussed in the next subsection.
With the Levi-Civita connection, (\ref{eq:EL1-1}) returns to the
original EFE. Therefore, by considering the projective transformation,
the Palatini formalism gives a similar dynamics with the second-order
formulation \cite{Iosifidis4}.

\subsubsection*{The Projective Invariance Problem}

The projective transformation is defined as follows \cite{Sotiriou,Iosifidis,Iosifidis4}:
\begin{equation}
\omega_{\mu\,\:\beta}^{\:\:\alpha}\rightarrow\omega_{\mu\,\:\beta}^{\:\:\alpha}+\delta_{\beta}^{\alpha}\xi_{\mu}.\label{eq:projectivetranfs}
\end{equation}
By a direct calculation, one could show that the Ricci scalar $\mathcal{R}$
is invariant under (\ref{eq:projectivetranfs}). This property could
be utilized to obtain the Levi-Civita condition from (\ref{eq:EL2})
as follows. First, one could show that the Palatini tensor is traceless:
$P_{\mu}^{\:\:\alpha\mu}=0$ \cite{Iosifidis,Iosifidis4}. Therefore,
in 4-dimension, one could only obtain 60 independent equations from
(\ref{eq:EL2}). These equations are not enough to determine completely
the connection, as it will need 64 independent equations. Instead,
the maximal condition on the connection one could obtain from (\ref{eq:EL2})
is:
\begin{equation}
\omega_{\mu\,\:\beta}^{\:\:\alpha}=\Gamma_{\mu\,\:\beta}^{\:\:\alpha}-\frac{1}{(n-1)}\underset{T_{\mu}}{\underbrace{T_{\lambda\,\:\mu}^{\:\:\lambda}}}\delta_{\beta}^{\alpha}=\Gamma_{\mu\,\:\beta}^{\:\:\alpha}+\frac{1}{n}\underset{Q_{\mu}}{\underbrace{\left(-\frac{1}{2}g^{\sigma\lambda}\nabla_{\mu}g_{\sigma\lambda}\right)}}\delta_{\beta}^{\alpha},\label{eq:solutionwithgauge}
\end{equation}
with $n$ is the dimension of $\mathcal{M}$, as shown in a detailed
derivation carried in \cite{Iosifidis,Iosifidis4}. $\Gamma_{\mu\,\:\beta}^{\:\:\alpha}$
is the Levi-Civita connection, while $T_{\mu}$ and $Q_{\mu},$ are
respectively, the torsion and disformation vectorial degrees of freedom
($Q_{\mu}$ is also known as the Weyl vector). Note that there exist
differences in the factors in front of $T_{\mu}$ and $Q_{\mu}$ with
\cite{Iosifidis,Iosifidis4}; these are due to the different definition
on torsion tensor (\ref{eq:torsion}). 

However, since the action (\ref{eq:Palatini}) is invariant under
(\ref{eq:projectivetranfs}), one could eliminate the vectorial degrees
of freedom using the transformation (\ref{eq:projectivetranfs}),
by setting an appropriate value of $\xi_{\mu}.$ In this article,
we only consider the case where $\xi_{\mu}=\frac{1}{3}T_{\mu}$, with
$n=4$. For the case where $\xi_{\mu}=-\frac{1}{n}Q_{\mu}$, or moreover,
a linear combination of $T_{\mu}$ and $Q_{\mu}$, one could consult
\cite{Iosifidis4,Smalley}. In the end, by performing the projective
transformation (\ref{eq:projectivetranfs}) on (\ref{eq:solutionwithgauge}),
one could fix the 4 vectorial degrees of freedom on $\omega_{\mu\,\:\beta}^{\:\:\alpha}$
such that (\ref{eq:solutionwithgauge}) becomes Levi-Civita:
\[
\omega_{\mu\,\:\beta}^{\:\:\alpha}=\Gamma_{\mu\,\:\beta}^{\:\:\alpha},
\]
with the condition:
\begin{equation}
T_{\mu}=T_{\lambda\,\:\mu}^{\:\:\lambda}=C_{\lambda\,\:\mu}^{\:\:\lambda}=0.\label{eq:projectcon}
\end{equation}
(\ref{eq:projectcon}) is known as the traceless torsion constraint.
It needs to be kept in mind that the tracelessness of the torsion
(\ref{eq:projectcon}) is introduced at the kinematical level; it
does not result from the dynamics, i.e. the equation of motion.

To obtain the Levi-Civita connection from the solution of the equation
of motion, one needs to add a new term corresponding to the constraint
into the action \cite{Sotiriou}:
\begin{equation}
S_{EH}\left[g,\omega,\chi\right]=\intop_{\mathcal{M}}\mathcal{R}\left[\omega\right]\mathrm{vol}\left[g\right]+S_{\mathrm{matter}}\left[g\right]+S_{LM}\left[\chi\right],\label{eq:PalatiniplusLM}
\end{equation}
where $S_{LM}=\intop_{\mathcal{M}}\chi^{\mu}T_{\mu}\textrm{vol }$and
$\chi^{\mu}$ is the Lagrange multiplier corresponding to $T_{\mu}$.
Minimizing (\ref{eq:PalatiniplusLM}) with respect to $g,$ $\omega,$
and $\chi$ gives:
\begin{align}
R_{\left(\alpha\beta\right)}-\frac{1}{2}\left(\mathcal{R}+\chi^{\mu}T_{\lambda\,\:\mu}^{\:\:\lambda}\right)g_{\alpha\beta} & =\mathcal{\kappa T}_{\alpha\beta},\label{eq:ELPalatini1}\\
\left(T_{\lambda\,\:\nu}^{\:\:\lambda}+Q_{\lambda\,\:\nu}^{\:\:\lambda}-\nabla_{\nu}\right)\left(2\delta_{\mu}^{\:\left[\nu\right.}g^{\left.\alpha\right]\beta}\right)+T_{\mu}^{\:\:\alpha\beta} & =\chi^{\alpha}\delta_{\mu}^{\beta}-\chi^{\beta}\delta_{\mu}^{\alpha},\label{eq:ELPalatini2}\\
T_{\lambda\,\:\mu}^{\:\:\lambda} & =0,\label{eq:ELPalatini3}
\end{align}
that could be solved to obtain $\chi^{\alpha}=0,$ hence, giving the
Levi-Cita condition and the standard EFE. One could conclude that
the Palatini formalism of gravity \textit{with} the traceless torsion
constraint is equivalent with GR \cite{Iosifidis,Iosifidis4}.

\subsubsection*{Metric-Affine Gravity, Metric-Affine $f\left(\mathcal{R}\right)$-Gravity,
and Metric-Affine GR}

Metric-Affine-Gravity (MAG) is a large class of theories based on
the first-order formalism with a general affine connection that includes
torsion and non-metricity. The choice of action for MAG could vary
greatly: the Ricci scalar $\mathcal{R}$ for the Metric-Affine General
Relativity (or Generalized Palatini), the power series of Ricci scalar
$f\left(\mathcal{R}\right)$ for the Metric-Affine $f\left(\mathcal{R}\right)$
Gravity, and other exotic actions such as $f\left(\mathcal{R},R_{\mu\nu}R^{\mu\nu}\right)$
and $\mathcal{L}\left(g_{\mu\nu},R_{\mu\nu\;\beta}^{\quad\alpha}\right)$
\cite{Iosifidis4}. In this article, we only consider Metric-Affine
$f\left(\mathcal{R}\right)$ Gravity and Metric-Affine General Relativity.

The action of Metric-Affine $f\left(\mathcal{R}\right)$-gravity is
defined as:
\begin{equation}
S\left[g,\omega,\chi\right]=\intop_{\mathcal{M}}f\left(\mathcal{R}\left[\omega\right]\right)\mathrm{vol}+S_{\mathrm{matter}}\left[g,\omega\right]+S_{\mathrm{LM}}\left[\chi\right].\label{eq:MAfR}
\end{equation}
Notice that now the matter action $S_{\mathrm{matter}}$ is also a
functional of the affine connection $\omega$. The variation of $S_{\mathrm{matter}}$
with respect to $\omega$ gives the hypermomentum tensor $\mathcal{H}$:
\begin{equation}
\delta_{\omega}S_{\mathrm{matter}}\left[g,\omega\right]=-\kappa\mathcal{H}_{\:\:\mu}^{\alpha\,\:\beta}\delta\omega_{\alpha\,\:\beta}^{\:\:\mu}\mathrm{vol},\label{eq:hypemom}
\end{equation}
and therefore, minimizing the action (\ref{eq:MAfR}) with respect
to the $g,$ $\omega,$ and $\chi$, results in three equations of
motion:
\begin{align}
f'\left(\mathcal{R}\right)R_{\left(\alpha\beta\right)}-\frac{1}{2}\left(f\left(\mathcal{R}\right)+\chi^{\mu}T_{\lambda\,\:\mu}^{\:\:\lambda}\right)g_{\alpha\beta} & =\mathcal{\kappa T}_{\alpha\beta},\label{eq:ELmafr1}\\
\left(\left(T_{\lambda\,\:\nu}^{\:\:\lambda}+Q_{\lambda\,\:\nu}^{\:\:\lambda}-\nabla_{\nu}\right)\left(2\delta_{\mu}^{\:\left[\nu\right.}g^{\left.\alpha\right]\beta}\right)+T_{\mu}^{\:\:\alpha\beta}\right)f'\left(\mathcal{R}\right) & =\kappa\mathcal{H}_{\:\:\mu}^{\alpha\,\:\beta}+2\chi^{\left[\alpha\right.}\delta_{\mu}^{\left.\beta\right]},\label{eq:ELmafr2}\\
T_{\lambda\,\:\mu}^{\:\:\lambda} & =0.\label{eq:ELmaffr3}
\end{align}
The last equation is the constraint equation. Notice that the hypermomentum
only exists if the matter term $S_{\mathrm{matter}}$ is a functional
of the connection, hence in the Palatini formalism, $\mathcal{H}=0$.
One could show that the torsion enters the dynamics through the antisymmetric
part of the second and third indices $\mathcal{H}^{\alpha\left[\mu\beta\right]},$
while the non-metricity enters through the symmetric part of the first
and third indices $\mathcal{H}^{\left(\alpha\right|\mu\left|\beta\right)},$
and if $\mathcal{H}_{\:\:\mu}^{\alpha\,\:\beta}=0,$ (\ref{eq:ELmafr2})
and (\ref{eq:ELmaffr3}) give the requirements of Levi-Civita connection,
hence the theory coincides with the original GR, for $f\left(\mathcal{R}\right)=\mathcal{R}$.
The equations of motions (\ref{eq:ELmafr2}) and (\ref{eq:ELmaffr3})
are the crucial results in the $f\left(\mathcal{R}\right)$-theory
of gravity. However, due to the scope of this article, our focus will
be on the stress-energy-momentum equation (\ref{eq:ELmafr1}).

The theory of Metric-Affine General Relativity (MAGR) or Generalized
Palatini theory could be obtained from Metric-Affine $f\left(\mathcal{R}\right)$-gravity
by setting $f\left(\mathcal{R}\right)=\mathcal{R}$. With this requirement,
(\ref{eq:ELmafr1})-(\ref{eq:ELmaffr3}) becomes:
\begin{align}
R_{\left(\alpha\beta\right)}-\frac{1}{2}\left(\mathcal{R}+\chi^{\mu}T_{\lambda\,\:\mu}^{\:\:\lambda}\right)g_{\alpha\beta} & =\mathcal{\kappa T}_{\alpha\beta},\label{eq:ELmagr1}\\
\left(T_{\lambda\,\:\nu}^{\:\:\lambda}+Q_{\lambda\,\:\nu}^{\:\:\lambda}-\nabla_{\nu}\right)\left(2\delta_{\mu}^{\:\left[\nu\right.}g^{\left.\alpha\right]\beta}\right)+T_{\mu}^{\:\:\alpha\beta} & =\kappa\mathcal{H}_{\:\:\mu}^{\alpha\,\:\beta}+2\chi^{\left[\alpha\right.}\delta_{\mu}^{\left.\beta\right]}\label{eq:ELmagr2}\\
T_{\lambda\,\:\mu}^{\:\:\lambda} & =0.\label{eq:ELmagr31}
\end{align}
which could be simplified to obtain:
\begin{align}
R_{\left(\alpha\beta\right)}-\frac{1}{2}\mathcal{R}g_{\alpha\beta} & =\mathcal{\kappa T}_{\alpha\beta},\label{eq:EoM1}\\
\left(Q_{\lambda\,\:\nu}^{\:\:\lambda}-\nabla_{\nu}\right)\left(2\delta_{\mu}^{\:\left[\nu\right.}g^{\left.\alpha\right]\beta}\right)+T_{\mu}^{\:\:\alpha\beta} & =\kappa\left(\mathcal{H}_{\:\:\mu}^{\alpha\,\:\beta}-\frac{2}{3}\mathcal{H}_{\:\:\;\;\,\sigma}^{\left[\alpha\right|\,\:\sigma}\delta_{\mu}^{\left|\beta\right]}\right),\label{eq:EoM2}\\
T_{\lambda\,\:\mu}^{\:\:\lambda} & =0,\label{eq:EoM3}
\end{align}
by inserting (\ref{eq:ELmagr31}) to (\ref{eq:ELmagr1})-(\ref{eq:ELmagr2})
and solving $\chi$ from (\ref{eq:ELmagr2}). Let us first focus only
on the stress-energy-momentum equation (\ref{eq:EoM1}). The quantity
in the LHS of (\ref{eq:EoM1}) is known as the generalized Einstein
tensor:
\[
G_{\alpha\beta}=R_{\left(\alpha\beta\right)}-\frac{1}{2}\mathcal{R}g_{\alpha\beta},
\]
which is symmetric on the $\left(\alpha,\beta\right)$-indices. The
stress-energy-momentum equation (\ref{eq:EoM1}) could be written
in an equivalent form as follows \cite{Eric}:
\begin{equation}
R_{\left(\alpha\beta\right)}=\kappa\mathcal{T}_{\alpha\beta}-\frac{1}{2}\mathcal{T}g_{\alpha\beta},\label{eq:efenice}
\end{equation}
with $\mathcal{T}=g^{\alpha\beta}\mathcal{T}_{\alpha\beta}$ is the
trace of the stress-energy momentum tensor (\ref{eq:tiga}). In the
next part of this section, (\ref{eq:EoM1}) will be decomposed into
its temporal and spatial part using the GCM equations derived in (\ref{eq:e}). 

\subsection{(3+1) Decomposition for MAGR}

\subsubsection*{The Adapted Coordinate, Lapse Function, and Shift Vector}

Let $\mathcal{M}$ be a globally hyperbolic Lorentzian manifold, and
let $x^{\mu}=\left\{ x^{0},x^{i}\right\} $ be a general local coordinate
on $\mathcal{M}.$ The coordinate vector basis on $T_{p}\mathcal{M}$
is $\partial_{\mu}=\left\{ \partial_{0},\partial_{i}\right\} $. Let
$\partial_{0}$ be the temporal component of $\partial_{\mu},$ that
could be decomposed according to the hypersurface $\Sigma$ as follows:
\begin{equation}
\partial_{0}=\underset{\boldsymbol{N}=N^{\mu}\partial_{\mu}}{\underbrace{\partial_{0}+g\left(\partial_{0},\hat{n}\right)\hat{n}}}+\underset{N}{\underbrace{-g\left(\partial_{0},\hat{n}\right)}\hat{n}},\label{eq:lapseshift}
\end{equation}
(and hence we take the lower part of the $\pm$ signature in (\ref{eq:lapseshift})).
The scalar $N$ is the lapse function describing the normal component
of $\partial_{0}$, while the vector $\boldsymbol{N}=N^{\mu}\partial_{\mu}$
is the shift vector describing the parallel part of $\partial_{0}$.

Let the metric $g$ be written in terms of the component in the local
coordinate $\partial_{\mu},$ using (\ref{eq:lapseshift}):
\begin{align}
g\left(\partial_{0},\partial_{0}\right) & =g_{00}=N^{\mu}N_{\mu}-N^{2},\label{eq:g1}\\
g\left(\partial_{0},\partial_{i}\right) & =g_{0i}=N_{i}+Nn_{i},\label{eq:g2}\\
g\left(\partial_{i},\partial_{j}\right) & =g_{ij}=\,^{3}q_{ij}-n_{i}n_{j}.\label{eq:g3}
\end{align}
Comparing with (\ref{eq:1a}), one could obtain $q_{00}=N_{0}=N^{\mu}N_{\mu},$
$q_{0i}=N_{i},$ and $n_{0}=-N$.

Let us consider the adapted coordinate on $\mathcal{M}$ (it had been
mentioned on the previous sections) where $\left\{ x^{i}\right\} $,
$i=1,2,3,$ is a local coordinate on $\Sigma$. Notice that in this
special coordinate, $n\perp\partial_{i}$, so that $n_{i}=g\left(n,\partial_{i}\right)=0$.
Moreover, in this coordinate, the shift $\boldsymbol{N}\in T_{p}\Sigma$
does not have a temporal component, namely $N^{0}=0.$ Using the adapted
coordinate, the spatial and the temporal part of $\Sigma$ could be
cleanly separated.

The components of metric $g$ could be written in the adapted coordinate
as follows; (\ref{eq:g1})-(\ref{eq:g3}) becomes:
\begin{align*}
g\left(\partial_{0},\partial_{0}\right) & =g_{00}=N^{i}N_{i}-N^{2},\\
g\left(\partial_{0},\partial_{i}\right) & =g_{0i}=N_{i},\\
g\left(\partial_{i},\partial_{j}\right) & =g_{ij}=\,^{3}q_{ij},
\end{align*}
while the components of the inverse metric $g^{-1},$ using the convention
in (\ref{eq:convention}), are:
\begin{align*}
g^{*}\left(dx^{0},dx^{0}\right) & =g^{00}=-N^{-2},\\
g^{*}\left(dx^{0},dx^{i}\right) & =g^{0i}=N^{i}N^{-2},\\
g^{*}\left(dx^{i},dx^{j}\right) & =g^{ij}=\,^{3}q^{ij}-\left(N^{i}N^{j}\right)N^{-2},
\end{align*}
with the coordinate basis vector on $T_{p}^{*}\mathcal{M}$ satisfies
$dx^{0}=-\hat{n}^{*}N^{-1}$ and $dx^{i}=\hat{n}^{*}N^{i}N^{-1}+\,^{3}dx^{i}$.
Notice that $dx^{i}$ is not necessarily equal to $\,^{3}dx^{i}=\,^{3}q^{ij}\,^{3}q\left(\partial_{j}\right)$
in (\ref{eq:d}).

In the adapted coordinate, one could check that the following relations
are true:
\begin{align*}
n & =n^{0}\partial_{0}+n^{i}\partial_{i}=N^{-1}\left(\partial_{0}-N^{i}\partial_{i}\right),\\
n^{*} & =n_{0}dx^{0}+n_{i}dx^{i}=-Ndx^{0},\\
\boldsymbol{N} & =N^{0}\partial_{0}+N^{i}\partial_{i}=-Nn^{i}\partial_{i},\\
\boldsymbol{N}^{*} & =N_{0}dx^{0}+N_{i}dx^{i}=N^{i}N_{i}dx^{0}+\,^{3}q_{ij}N^{j}dx^{i}.
\end{align*}
These relations will be useful for the following derivation.

Applying the adapted coordinate to the Riemann curvature and torsion
decomposition, namely (\ref{eq:e}) and (\ref{eq:torsion1})-(\ref{eq:torsion2}),
one obtains:
\begin{align}
R_{ij\,\:\,k}^{\:\:\:\:0}= & \left(\,^{3}\nabla_{i}K_{jk}-\,^{3}\nabla_{j}K_{ik}+\,^{3}T_{i\,\:\,j}^{\:\:l}K_{lk}+\Theta_{i}K_{jk}-\Theta_{j}K_{ik}\right)n^{0},\label{eq:gcm1}\\
R_{ij\,\:\,l}^{\:\:\:\:k}= & \,^{3}R_{ij\,\:\,l}^{\:\:\:\:k}+K_{jl}\mathcal{K}_{i}^{\:\:k}-K_{il}\mathcal{K}_{j}^{\:\:k}+\underset{\frac{1}{n^{0}}R_{ij\,\:\,l}^{\:\:\:\:\:0}}{\underbrace{\left(\,^{3}\nabla_{i}K_{jl}-\,^{3}\nabla_{j}K_{il}+\,^{3}T_{i\,\:\:\:j}^{\,\:m}K_{ml}+\Theta_{i}K_{jl}-\Theta_{j}K_{il}\right)}}n^{k},\label{eq:gcm2}
\end{align}
which are, respectively, the components of Codazzi-Mainardi and Gauss
equations, and:
\begin{eqnarray}
\hat{n}_{\mu}T_{i\,\:\:j}^{\:\:\mu} & = & K_{ij}-K_{ji},\label{eq:T1}\\
g_{k\mu}T_{i\,\:\:j}^{\:\:\mu} & = & \,^{3}q_{kl}\,^{3}T_{i\,\:\,j}^{\:\:l},\label{eq:T2}
\end{eqnarray}
which are the components of torsion decomposition. Here, the indices
of the extrinsic curvatures are raised with the 3-metric, for example,
$\mathcal{K}_{i}^{\:\:j}=\,^{3}q^{jk}\,^{3}\mathcal{K}_{ik}$.

\subsubsection*{The Energy Part}

Let us consider the purely-time part of the Einstein tensor, where
the $\alpha,\beta$ indices of $G_{\alpha\beta}$ is contracted with
the normal $\hat{n}$ to give the following scalar quantity:
\begin{equation}
G_{\alpha\beta}n^{\alpha}n^{\beta}=G\left(\hat{n},\hat{n}\right)=\mathbf{Ric}\left(\hat{n},\hat{n}\right)-\frac{1}{2}g\left(\hat{n},\hat{n}\right)\mathcal{R}.\label{eq:ham1}
\end{equation}
By a direct calculation, one could show that the generalized Ricci
scalar could be decomposed into:
\begin{equation}
\mathcal{R}=\,^{3}\mathcal{R}-\mathrm{tr}\left(K\mathcal{K}\right)+\left(\mathrm{tr}K\right)\left(\mathrm{tr}\mathcal{K}\right)-\mathrm{\mathbf{Ric}}\left(\hat{n},\hat{n}\right)+\overline{\mathrm{\mathbf{Ric}}}\left(\hat{n},\hat{n}\right),\label{eq:Riccidec}
\end{equation}
where $\overline{\mathrm{\mathbf{Ric}}}$ is the co-Ricci tensor satisfying
(\ref{eq:coRic}). Inserting (\ref{eq:Riccidec}) to (\ref{eq:ham1})
gives:

\[
G\left(\hat{n},\hat{n}\right)=\frac{1}{2}\left(\mathrm{\mathbf{Ric}}\left(\hat{n},\hat{n}\right)+\overline{\mathrm{\mathbf{Ric}}}\left(\hat{n},\hat{n}\right)\right)+\frac{1}{2}\left(\,^{3}\mathcal{R}-\mathrm{tr}\left(K\mathcal{K}\right)+\left(\mathrm{tr}K\right)\left(\mathrm{tr}\mathcal{K}\right)\right).
\]
The first term is the additional part due to the non-metricity and
torsion, this will be clear in the next subsections.

The normal part of the stress-energy-momentum tensor (\ref{eq:tiga})
is the energy density, namely $\mathcal{T}\left(\hat{n},\hat{n}\right)=E,$
and the energy equation in MAG is:
\begin{equation}
\frac{1}{2}\left(\mathrm{\mathbf{Ric}}\left(\hat{n},\hat{n}\right)+\overline{\mathrm{\mathbf{Ric}}}\left(\hat{n},\hat{n}\right)\right)+\frac{1}{2}\left(\,^{3}\mathcal{R}-\mathrm{tr}\left(K\mathcal{K}\right)+\left(\mathrm{tr}K\right)\left(\mathrm{tr}\mathcal{K}\right)\right)=\kappa E.\label{eq:Hamconstraint}
\end{equation}
For metric connections, $\mathrm{\mathbf{Ric}}=-\overline{\mathrm{\mathbf{Ric}}}$
and $K=\mathcal{K},$ hence, for Levi-Civita connection, (\ref{eq:Hamconstraint})
returns to the original form, the Hamiltonian constraint. Notice that
(\ref{eq:Hamconstraint}) is also valid for a metric connection with
torsion, since the antisymmetric part of $\,^{3}\mathcal{R}$ and
$K$, resulting from a non-vanishing torsion, do not contribute to
the energy equation.

\subsubsection*{The Momentum Part}

The momentum part of the Einstein tensor is a mixture between the
time and spatial parts. In the adapted coordinate, it could be written
as follows:
\begin{align}
G_{i\mu}n^{\mu}=G\left(\partial_{i},\hat{n}\right)=G\left(\hat{n},\partial_{i}\right)= & \frac{1}{2}\left(\mathrm{\mathbf{Ric}}\left(\partial_{i},\hat{n}\right)+\mathrm{\mathbf{Ric}}\left(\hat{n},\partial_{i}\right)\right).\label{eq:dff}
\end{align}
where the term containing the Ricci scalar is zero due to the fact
that in the adapted coordinate, $g\left(\partial_{i},\hat{n}\right)=n_{i}=0.$
From a direct calculation, one could show that the terms containing
the 3-covariant derivative of the extrinsic curvature in (\ref{eq:gcm1})
comes from the co-Ricci tensor, instead of the Ricci tensor:
\begin{equation}
\mathrm{\overline{\mathbf{Ric}}}\left(\partial_{i},\hat{n}\right)=q^{jk}\left(\,^{3}\nabla_{i}K_{jk}-\,^{3}\nabla_{j}K_{ik}+\,^{3}T_{i\,\:\,j}^{\:\:l}K_{lk}+\Theta_{i}K_{jk}-\Theta_{j}K_{ik}\right)-g\left(\hat{n},\boldsymbol{R}\left(\hat{n},\partial_{i}\right)\hat{n}\right).\label{eq:aneeh}
\end{equation}
For a metric connection, $\mathrm{\mathbf{Ric}}=-\overline{\mathrm{\mathbf{Ric}}}$,
so one could immediately insert (\ref{eq:aneeh}) to (\ref{eq:dff}).
This is not the case for a general connection. Hence, in MAGR, there
is no direct way to write the momentum equation in terms of the intrinsic
curvature of the first kind $K$. Let us postponed this problem for
the moment and write the momentum equation as follows:
\begin{align}
\frac{1}{2}\left(\mathrm{\mathbf{Ric}}\left(\partial_{i},\hat{n}\right)+\mathrm{\mathbf{Ric}}\left(\hat{n},\partial_{i}\right)\right) & =\kappa p_{i},\label{eq:diffeom}
\end{align}
with $p_{i}$ is the 'mixed' part of the stress-energy-momentum tensor
(\ref{eq:tiga}), namely, the (relativistic) momentum $\mathcal{T}\left(\partial_{i},\hat{n}\right)=\mathcal{T}\left(\hat{n},\partial_{i}\right)=p_{i}.$

For a metric connection with torsion, the momentum equation becomes:
\begin{equation}
\,^{3}\nabla_{j}K_{i}^{\:\:j}-\,^{3}\nabla_{i}K_{j}^{\:\:j}-\,^{3}T_{i\,\:\,j}^{\:\:k}K_{k}^{\:\:j}+\frac{1}{2}n^{\nu}\left(\nabla_{\mu}T_{\nu\,\:i}^{\:\:\mu}\right)=\kappa p_{i},\label{eq:torsiondiff}
\end{equation}
where in general, the extrinsic curvature $K$ contains an antisymmetric
part. For the Levi-Civita case, the torsion vanishes, and (\ref{eq:diffeom})
returns to the original momentum (or diffeomorphism) constraint, with
symmetric $K$.

\subsubsection*{The Stress-Energy Part}

The purely-spatial part of the stress-energy-momentum equation are
the following set of equations:
\begin{equation}
G_{ij}=G\left(\partial_{i},\partial_{j}\right)=\frac{1}{2}\left(\mathbf{Ric}\left(\partial_{i},\partial_{j}\right)+\mathbf{Ric}\left(\partial_{j},\partial_{i}\right)\right)-\frac{1}{2}g\left(\partial_{i},\partial_{j}\right)\mathcal{R}.\label{eq:dyns2}
\end{equation}
The first term, namely, the spatial part of the Ricci tensor, could
be decomposed as follows:
\begin{equation}
\mathbf{Ric}\left(\partial_{i},\partial_{j}\right)=\,^{3}\mathbf{Ric}\left(\partial_{i},\partial_{j}\right)+K_{ij}\mathrm{tr}\mathcal{K}-\mathcal{K}_{i}^{\:\:k}K_{kj}-g\left(\hat{n},\boldsymbol{R}\left(\hat{n},\partial_{i}\right)\partial_{j}\right).\label{eq:ricdec}
\end{equation}
Inserting (\ref{eq:ricdec}) and the Ricci scalar (\ref{eq:Riccidec})
to (\ref{eq:dyns2}) gives immediately the spatial part of $G$:
\begin{align*}
G\left(\partial_{i},\partial_{j}\right)= & \,^{3}G\left(\partial_{i},\partial_{j}\right)-g\left(\hat{n},\boldsymbol{R}\left(\hat{n},\partial_{\left(i\right.}\right)\partial_{\left.j\right)}\right)+K_{\left(ij\right)}\mathrm{tr}\mathcal{K}-\mathcal{K}_{\left(i\right|}^{\,\:\:k}K_{k\left|j\right)}\\
 & \qquad\qquad\quad-\frac{1}{2}\,^{3}q_{ij}\left(\left(\mathrm{tr}K\right)\left(\mathrm{tr}\mathcal{K}\right)-\mathrm{tr}\left(K\mathcal{K}\right)-\mathbf{Ric}\left(\hat{n},\hat{n}\right)+\overline{\mathbf{Ric}}\left(\hat{n},\hat{n}\right)\right),
\end{align*}
with:
\[
\,^{3}G\left(\partial_{i},\partial_{j}\right)=\frac{1}{2}\left(\,^{3}\mathbf{Ric}\left(\partial_{i},\partial_{j}\right)+\,^{3}\mathbf{Ric}\left(\partial_{j},\partial_{i}\right)-\,^{3}q_{ij}\,^{3}\mathcal{R}\right)
\]
is the Einstein tensor on $\Sigma.$

The spatial part of the stress-energy-momentum tensor (\ref{eq:tiga})
is the stress tensor $\mathcal{T}\left(\partial_{i},\partial_{j}\right)=\mathcal{S}\left(\partial_{i},\partial_{j}\right)=\mathcal{S}\left(\partial_{j},\partial_{i}\right),$
hence the stress-energy equations of EFE is:
\begin{align}
\,^{3}G\left(\partial_{i},\partial_{j}\right)-g\left(\hat{n},\boldsymbol{R}\left(\hat{n},\partial_{\left(i\right.}\right)\partial_{\left.j\right)}\right)+K_{\left(ij\right)}\mathrm{tr}\mathcal{K}-\mathcal{K}_{\left(i\right|}^{\,\:\:k}K_{k\left|j\right)}\qquad\qquad\quad\label{eq:stress}\\
-\frac{1}{2}\,^{3}q_{ij}\left(\left(\mathrm{tr}K\right)\left(\mathrm{tr}\mathcal{K}\right)-\mathrm{tr}\left(K\mathcal{K}\right)-\mathbf{Ric}\left(\hat{n},\hat{n}\right)+\overline{\mathbf{Ric}}\left(\hat{n},\hat{n}\right)\right) & =\kappa\mathcal{S}\left(\partial_{i},\partial_{j}\right).\nonumber 
\end{align}
One could write (\ref{eq:stress}) in a more convenient way as follows.
Projecting (\ref{eq:efenice}) to the hypersurface $\Sigma$ gives:
\[
\frac{1}{2}\left(\mathbf{Ric}\left(\partial_{i},\partial_{j}\right)+\mathbf{Ric}\left(\partial_{j},\partial_{i}\right)\right)=\kappa\mathcal{S}\left(\partial_{i},\partial_{j}\right)-\frac{1}{2}g\left(\partial_{i},\partial_{j}\right)\kappa\mathcal{T}.
\]
Using (\ref{eq:ricdec}), one could obtain:
\begin{equation}
\,^{3}R_{\left(ij\right)}+K_{\left(ij\right)}\mathrm{tr}\mathcal{K}-\mathcal{K}_{\left(i\right|}^{\,\:\:k}K_{k\left|j\right)}-g\left(\hat{n},\boldsymbol{R}\left(\hat{n},\partial_{\left(i\right.}\right)\partial_{\left.j\right)}\right)=\kappa\mathcal{S}_{ij}-\frac{1}{2}\,^{3}q_{ij}\kappa\mathcal{T}.\label{eq:stress2}
\end{equation}

In the original GR, the purely-spatial part of the EFE is the only
part that contains the dynamics of the system, i.e., the equations
which contain the time derivative of the 3-metric. In (\ref{eq:stress2}),
the time derivative is hidden such that it is contained implicitly
in the covariant term $g\left(\hat{n},\boldsymbol{R}\left(\hat{n},\partial_{\left(i\right.}\right)\partial_{\left.j\right)}\right).$
We will show that this is indeed the case in the following subsections.

\subsubsection*{The Additional Variables on the Hypersurface}

In the geometrodynamics concept introduced by Wheeler \cite{Wheeler},
the system of GR could be equivalently described using only fields
in $\Sigma$ which evolve in time. In this perspective, one does not
need to refer to the 4-manifold $\mathcal{M}$ explicitly. However,
the EFE (\ref{eq:Hamconstraint}), (\ref{eq:diffeom}), and (\ref{eq:stress2})
contain some terms that are still covariant, therefore, these terms
need to be decomposed into the temporal and spatial parts. 

Let us regard $\hat{n}$, the unit normal to $\Sigma$, as a 4-velocity
in a Eulerian frame. The 4-acceleration, defined as $\nabla_{\hat{n}}\hat{n}=\alpha=\alpha^{\mu}\partial_{\mu}$
in (\ref{eq:accell}), has time and spatial components as follows:
\begin{align}
\left\langle \hat{n}^{*},\nabla_{\hat{n}}\hat{n}\right\rangle  & =g\left(\nabla_{\hat{n}}\hat{n},\hat{n}\right)=g\left(\alpha,\hat{n}\right):=-\Theta\left(\hat{n}\right),\label{eq:1a-1}\\
\left\langle \,^{3}dx^{i},\nabla_{\hat{n}}\hat{n}\right\rangle  & =\left\langle \,^{3}dx^{i},\alpha\right\rangle :=\alpha^{i}.\label{eq:2a}
\end{align}
Equation (\ref{eq:1a-1}), already defined in (\ref{eq:angle}), is
the angle between the 4-velocity with the 4-acceleration which is
zero for a metric connection. (\ref{eq:2a}) is the components of
the (relativistic) 3-acceleration, $\,^{3}\alpha=\alpha^{i}\partial_{i}$
(not to be confused with $\boldsymbol{a}_{i}=\nabla_{i}\hat{n}$ in
(\ref{eq:accell})).

The acceleration $\alpha$ is a part of the acceleration tensor $\boldsymbol{a}$
defined in (\ref{eq:accell}), the other part of the tensor are the
following:
\begin{align}
\left\langle \hat{n}^{*},\nabla_{i}\hat{n}\right\rangle =g\left(\nabla_{i}\hat{n},\hat{n}\right) & =g\left(\boldsymbol{a}_{i},\hat{n}\right):=-\Theta_{i},\label{eq:3a}
\end{align}
$\boldsymbol{a}_{i}$, together with $\boldsymbol{a}_{n}=\alpha$,
define the acceleration tensor $\boldsymbol{a}=\left(\boldsymbol{a}_{n},\boldsymbol{a}_{i}\right).$
$\boldsymbol{a}_{i}$ is the rate of change of the 4-velocity in the
spatial direction $\partial_{i}$; therefore $\Theta_{i}$ is the
temporal components of $\boldsymbol{a}_{i}$ or the angle between
$\alpha_{i}$ and $\hat{n}$. One could similarly obtain the spatial
components of $\boldsymbol{a}_{i}$, but one could prove using equation
(\ref{eq:1a}), (\ref{eq:extrinsic2}), and (\ref{eq:pentingx}),
that this is only the extrinsic curvature of the second kind:
\begin{equation}
\left\langle \,^{3}dx^{j},\nabla_{i}\hat{n}\right\rangle =\,^{3}q^{jk}g\left(\nabla_{i}\hat{n},\partial_{k}\right)=\,^{3}q^{jk}g\left(\boldsymbol{a}_{i},\partial_{k}\right)=\,^{3}q^{jk}\mathcal{K}_{ik}=\mathcal{K}_{i}^{\:\:j},\label{eq:4a}
\end{equation}
which had been defined in (\ref{eq:extrinsic2}).

Having defined all the covariant derivative of $\hat{n}$ in all directions
and its components, now we could proceed to define the covariant derivative
of the spatial direction. The first two are the components of $\nabla_{i}\partial_{j}$,
the rate of change of $\partial_{j}$ in the direction $\partial_{i}$:
\begin{align}
\left\langle \hat{n}^{*},\nabla_{i}\partial_{j}\right\rangle  & =g\left(\nabla_{i}\partial_{j},\hat{n}\right)=g\left(\omega_{i\:\:\:j}^{\,\,\:\mu}\partial_{\mu},\hat{n}\right)=-K_{ij},\label{eq:5a}\\
\left\langle \,^{3}dx^{k},\nabla_{i}\partial_{j}\right\rangle  & =\left\langle \,^{3}dx^{k},\omega_{i\:\:\:j}^{\,\,\:\mu}\partial_{\mu}\right\rangle =\,^{3}\omega_{i\:\,\:j}^{\:\:k}.\label{eq:6a}
\end{align}
(\ref{eq:5a}) is the time component of $\nabla_{i}\partial_{j}$
which is exactly the extrinsic curvature of the first kind defined
in (\ref{eq:3}). The spatial component of $\nabla_{i}\partial_{j}$
is exactly the induced 3-connection (\ref{eq:6a}) on $\Sigma$.

The last quantities are the components of $\nabla_{\hat{n}}\partial_{i}$,
which define the rate of change of $\partial_{i}$ in the direction
of $\hat{n}$, i.e., the evolution of $\partial_{i}$ in time:
\begin{align}
\left\langle \hat{n}^{*},\nabla_{\hat{n}}\partial_{i}\right\rangle  & =g\left(\nabla_{\hat{n}}\partial_{i},\hat{n}\right)=g\left(n^{\alpha}\omega_{\alpha\:\,\:i}^{\,\:\:\mu}\partial_{\mu},\hat{n}\right)=\omega\left(\hat{n}\right)_{\;\:i}^{\mu}n_{\mu}:=-\Delta_{i},\label{eq:7a}\\
\left\langle \,^{3}dx^{j},\nabla_{\hat{n}}\partial_{i}\right\rangle  & =\left\langle \,^{3}dx^{j},n^{\alpha}\omega_{\alpha\:\,\:i}^{\,\:\:\mu}\partial_{\mu}\right\rangle =\omega\left(\hat{n}\right)_{\;\:i}^{j}:=\Delta_{\;\:i}^{j}.\label{eq:8a}
\end{align}
One could think of $\Delta_{i}$ and $\Delta_{\;\:i}^{j}$ as the
generator of the evolution of $\partial_{i}$. (\ref{eq:7a}) is the
temporal component of the evolution, which drags $\partial_{i}$ along
the normal direction, while (\ref{eq:8a}) are the spatial components
of the evolution, which moves $\partial_{i}$ along $\Sigma.$ Since
$\nabla$ is not only a differentiation $\partial$, but it also rotates
and shears vectors by $\omega$, the existence of the spatial components
(\ref{eq:8a}) is understandable.

Notice that in the original GR, the EFE could be written as functions
of the following variables: the extrinsic curvature, the 3-connection
(which is a function of metric) in terms of 3-Ricci scalar and 3-Einstein
tensor, the 3-acceleration in terms of the lapse $N$, and the evolution
part $\Delta_{\;\:i}^{j}$, which contains the shift $\boldsymbol{N}$.
These variables are not independent of one another. For MAGR (and
hence MAG), we have 8 different (with 4 additional) variables: $K,$
$\mathcal{K}$, $\,^{3}\alpha$, $\,^{3}\omega,$ $\Theta\left(\hat{n}\right),$
$\Theta_{i}$, $\Delta_{i}$, and $\Delta_{\;\:i}^{j}$.

With the full variables on $\Sigma$, one could write the decompositions
of the derivatives of $\hat{n}$ and $\partial_{i}$:
\begin{align}
\nabla_{\hat{n}}\hat{n} & =\Theta\left(\hat{n}\right)\hat{n}+\alpha^{i}\partial_{i},\label{eq:aa}\\
\nabla_{i}\hat{n} & =\Theta_{i}\hat{n}+\mathcal{K}_{i}^{\:\:j}\partial_{j},\label{eq:bb}\\
\nabla_{i}\partial_{j} & =K_{ij}\hat{n}+\,^{3}\omega_{i\:\,\:j}^{\:\:k}\partial_{k},\label{eq:cc}\\
\nabla_{\hat{n}}\partial_{i} & =\Delta_{i}\hat{n}+\Delta_{\;\:i}^{j}\partial_{j},\label{eq:dd}
\end{align}
to rederive the torsion and the non-metricity factor in terms of the
additional variables as follows. The decomposition of the torsion
tensor is:
\begin{align}
T\left(n,n\right) & =0,\label{eq:t1}\\
T\left(n,\partial_{i}\right)=-T\left(\partial_{i},n\right) & =\left(\Delta_{i}-\Theta_{i}-g\left(\partial_{i}\hat{n},\hat{n}\right)\right)\hat{n}+\left(\Delta_{\;\:i}^{j}-\mathcal{K}_{i}^{\:\:j}+\left\langle \,^{3}dx^{j},\partial_{i}\hat{n}\right\rangle \right)\partial_{j},\label{eq:t2}\\
T\left(\partial_{i},\partial_{j}\right) & =\,^{3}T\left(\partial_{i},\partial_{j}\right)+\left(K\left(\partial_{i},\partial_{j}\right)-K\left(\partial_{j},\partial_{i}\right)\right)\hat{n},\label{eq:t3}
\end{align}
while the decomposition of the non-metricity factor is:
\begin{align}
\nabla_{n}g^{*} & =-2\Theta\left(\hat{n}\right)\hat{n}\otimes\hat{n}+\left(\Delta^{i}-\alpha^{i}\right)\left(\hat{n}\otimes\partial_{i}+\partial_{i}\otimes\hat{n}\right)+\,^{3}\nabla_{n}\,^{3}q^{*}\label{eq:q1}\\
\nabla_{i}g^{*} & =-2\Theta_{i}\hat{n}\otimes\hat{n}+\left(K_{i}^{\;j}-\mathcal{K}_{i}^{\:\:j}\right)\left(\hat{n}\otimes\partial_{j}+\partial_{j}\otimes\hat{n}\right)+\,^{3}\nabla_{i}\,^{3}q^{*}\label{eq:q2}
\end{align}
where:
\begin{align}
\,^{3}\nabla_{n}\,^{3}q^{*}=\left(\,^{3}\nabla_{n}\,^{3}q^{ij}\right)\partial_{i}\otimes\partial_{j}= & \left(n\left[\,^{3}q^{ij}\right]+\Delta_{\;\:k}^{i}\,^{3}q^{kj}+\Delta_{\;\:k}^{j}\,^{3}q^{ki}\right)\partial_{i}\otimes\partial_{j},\label{eq:q3}\\
\,^{3}\nabla_{i}\,^{3}q^{*}=\left(\,^{3}\nabla_{i}\,^{3}q^{jk}\right)\partial_{i}\otimes\partial_{k}= & \left(\partial_{i}\,^{3}q^{jk}+\omega_{i}^{\;jk}+\omega_{i}^{\;kj}\right)\partial_{i}\otimes\partial_{k}.\label{eq:q4}
\end{align}
Therefore, the torsionless condition $T=0$ is equivalent to:
\begin{equation}
\begin{array}{ccccc}
\,^{3}\omega_{i\:\,\:j}^{\:\:k}=\,^{3}\omega_{j\:\,\:i}^{\:\:k}, & \qquad & \Delta_{i} & = & g\left(\partial_{i}\hat{n},\hat{n}\right)+\Theta_{i},\\
K_{ij}=K_{ji}, &  & \Delta_{\;\:i}^{j} & = & \mathcal{K}_{i}^{\:\:j}-\left\langle \,^{3}dx^{j},\partial_{i}\hat{n}\right\rangle ,
\end{array}\label{eq:tor1}
\end{equation}
with: 
\begin{align}
g\left(\hat{n},\partial_{i}\hat{n}\right) & =\partial_{i}\ln N,\label{eq:N}\\
\left\langle dx^{j},\partial_{i}\hat{n}\right\rangle  & =\frac{1}{N}\left(N^{j}\partial_{i}\ln N-\partial_{i}N^{j}\right),\label{eq:Nj}\\
\left\langle \,^{3}dx^{j},\partial_{i}\hat{n}\right\rangle  & =-\frac{1}{N}\partial_{i}N^{j},
\end{align}
while the metric compatibility $\nabla_{\mu}g^{*}=0$ is equivalent
to:
\begin{equation}
\begin{array}{cccc}
\Theta\left(\hat{n}\right) & =0, & \qquad & \,^{3}\alpha_{i}=\Delta_{i},\\
\Theta_{i} & =0, &  & \mathcal{K}_{ij}=K_{ij},
\end{array}\label{eq:met1}
\end{equation}
together with:
\begin{align}
\,^{3}\nabla_{n}\,^{3}q^{*}= & 0,\label{eq:met}\\
\,^{3}\nabla_{i}\,^{3}q^{*}= & 0.\nonumber 
\end{align}

The relation (\ref{eq:t1})-(\ref{eq:q2}) could be used to confirm
the theorems we proved in Section III: The metricity condition $\nabla_{\mu}g=0$
will cause $\,^{3}\nabla_{i}\,^{3}q^{*}=0,$ and $T\left(\partial_{\mu},\partial_{\nu}\right)=0$
will cause $\,^{3}T\left(\partial_{i},\partial_{j}\right),$ but these
relations are not valid vice-versa. The Levi-Civita connection must
satisfy the metric compatibility and torsionless condition, and hence,
the 8 additional variables are constrained by (\ref{eq:tor1})-(\ref{eq:met1}).
Applying these constraints, the variables in the stress-energy-momentum
equation  for the Levi-Civita connection reduce to 4: $K=\mathcal{K}$,
$\,^{3}\omega,$ $\,^{3}\alpha_{i}=\Delta_{i}=\partial_{i}\ln N$,
and $\Delta_{\;\:i}^{j}=\mathcal{K}_{i}^{\:\:j}+\frac{1}{N}\partial_{i}N^{j}$. 

\subsubsection*{The Results}

Relations (\ref{eq:aa})-(\ref{eq:dd}) are used to split the covariant
parts in (\ref{eq:Hamconstraint}), (\ref{eq:diffeom}), and (\ref{eq:stress2})
into (3+1) forms, in particular:
\begin{align*}
\mathrm{\mathbf{Ric}}\left(\hat{n},\hat{n}\right)= & \Theta\left(\hat{n}\right)\mathrm{tr}\mathcal{K}-\Theta_{i}\alpha^{i}-\hat{n}\left[\mathrm{tr}\mathcal{K}\right]-\mathcal{K}_{i}^{\:\:j}\Delta_{\;\:j}^{i}+\,^{3}\nabla_{i}\alpha^{i}+\alpha^{i}g\left(\hat{n},\partial_{i}\hat{n}\right)-\mathcal{K}_{j}^{\:\:i}\left\langle ^{3}dx^{j},\partial_{i}\hat{n}\right\rangle ,\\
\overline{\mathrm{\mathbf{Ric}}}\left(\hat{n},\hat{n}\right)= & \,^{3}q^{ij}\left(\Theta\left(\hat{n}\right)K_{ij}-\Theta_{i}\Delta_{j}+\hat{n}\left[K_{ij}\right]-K_{ik}\Delta_{\;\:j}^{k}-\,^{3}\nabla_{i}\Delta_{j}-\Delta_{j}g\left(\hat{n},\partial_{i}\hat{n}\right)+K_{kj}\left\langle ^{3}dx^{k},\partial_{i}\hat{n}\right\rangle \right),
\end{align*}
\begin{align*}
\mathrm{\mathbf{Ric}}\left(\hat{n},\partial_{i}\right)= & -\alpha^{j}K_{ji}-\hat{n}\left[\,^{3}\omega_{j\:\,\:i}^{\:\:j}\right]+\Delta_{i}\mathrm{tr}\mathcal{K}+\,^{3}\nabla_{j}\Delta_{\:\;i}^{j}+\Delta_{\:\;i}^{j}g\left(\hat{n},\partial_{j}\hat{n}\right)-\,^{3}\omega_{k\:\,\:i}^{\:\:j}\left\langle ^{3}dx^{k},\partial_{j}\hat{n}\right\rangle ,\\
\mathrm{\mathbf{Ric}}\left(\partial_{i},\hat{n}\right)= & \,^{3}\nabla_{j}\mathcal{K}_{i}^{\:\:j}-\,^{3}\nabla_{i}\mathcal{K}_{j}^{\:\:j}+\,^{3}T_{j\,\:\,i}^{\:\:k}\mathcal{K}_{k}^{\:\:j}+\Theta_{i}\mathcal{K}_{j}^{\:\:j}-\Theta_{j}\mathcal{K}_{i}^{\:\:j}\\
 & +\mathcal{K}_{i}^{\:\:j}\Delta_{j}-K_{ij}\alpha^{j}-\partial_{i}\Theta\left(\hat{n}\right)+\hat{n}\left[\Theta_{i}\right]-\Theta\left(\hat{n}\right)g\left(\hat{n},\partial_{i}\hat{n}\right)+\Theta_{j}\left\langle ^{3}dx^{j},\partial_{i}\hat{n}\right\rangle ,
\end{align*}
and:
\begin{align*}
g\left(\hat{n},\boldsymbol{R}\left(\hat{n},\partial_{i}\right)\partial_{j}\right)= & \,^{3}\nabla_{i}\Delta_{j}+\Theta_{i}\Delta_{j}-\Theta\left(\hat{n}\right)K_{ij}+K_{ik}\Delta_{\;\:j}^{k}-\hat{n}\left[K_{ij}\right]+\Delta_{j}g\left(\hat{n},\partial_{i}\hat{n}\right)-K_{kj}\left\langle \,^{3}dx^{k},\partial_{i}\hat{n}\right\rangle .
\end{align*}
One could observe that the Ricci tensor $\mathrm{\mathbf{Ric}}\left(\partial_{i},\hat{n}\right)$
contains the 3-covariant derivative of $\mathcal{K}$ instead of $K$
as in (\ref{eq:gcm1}). Together with (\ref{eq:N}) and (\ref{eq:Nj}),
the (3+1) field equations for MAG could be written in partial differential
equations containing only 3-dimensional variables on the hypersurface
$\Sigma$:
\begin{align}
 & \frac{1}{2}\left(\,^{3}\mathcal{R}-\mathrm{tr}\left(K\mathcal{K}\right)+\left(\mathrm{tr}K\right)\left(\mathrm{tr}\mathcal{K}\right)-\left(\mathcal{K}_{i}^{\:\:j}+K_{\:\:i}^{j}\right)\Delta_{\;\:j}^{i}+\,^{3}\nabla_{i}\alpha^{i}-\,^{3}q^{ij}\,^{3}\nabla_{i}\Delta_{j}+\,^{3}q^{ij}\hat{n}\left[K_{ij}\right]\right.\label{eq:satu}\\
 & \left.\quad-\hat{n}\left[\mathrm{tr}\mathcal{K}\right]-\Theta_{i}\left(\alpha^{i}+\,^{3}q^{ij}\Delta_{j}\right)+\Theta\left(\hat{n}\right)\left(\mathrm{tr}\mathcal{K}+\mathrm{tr}K\right)+\left(\alpha^{i}-\,^{3}q^{ij}\Delta_{j}\right)\partial_{i}\ln N+\frac{1}{N}\left(\mathcal{K}_{j}^{\:\:i}-K_{j}^{\,\:i}\right)\partial_{i}N^{j}\right)=\kappa E.\nonumber 
\end{align}
\begin{align}
 & \frac{1}{2}\left(\,^{3}\nabla_{j}\mathcal{K}_{i}^{\:\:j}-\,^{3}\nabla_{i}\mathcal{K}_{j}^{\:\:j}+\,^{3}T_{j\,\:\,i}^{\:\:k}\mathcal{K}_{k}^{\:\:j}-\alpha^{j}\left(K_{ij}+K_{ji}\right)+\,^{3}\nabla_{j}\Delta_{\;i}^{j}-\hat{n}\left[\,^{3}\omega_{j\:\,\:i}^{\:\:j}\right]+\hat{n}\left[\Theta_{i}\right]-\partial_{i}\Theta\left(\hat{n}\right)\right.\label{eq:dua}\\
 & \quad\left.+\left(\Delta_{i}+\Theta_{i}\right)\mathcal{K}_{j}^{\:\:j}+\left(\Delta_{j}-\Theta_{j}\right)\mathcal{K}_{i}^{\:\:j}+\Delta_{\;\:i}^{j}\partial_{j}\ln N-\Theta\left(\hat{n}\right)\partial_{i}\ln N+\frac{1}{N}\left(\,^{3}\omega_{k\:\,\:i}^{\:\:j}\partial_{j}N^{k}-\Theta_{j}\partial_{i}N^{j}\right)\right)=\kappa p_{i},\nonumber 
\end{align}
\begin{align}
 & \hat{n}\left[K_{\left(ij\right)}\right]-\frac{1}{N}K_{k\left(i\right.}\partial_{\left.j\right)}N^{k}-\left(\partial_{\left(i\right|}\ln N+\Theta_{\left(i\right|}-\,^{3}\nabla_{\left(i\right|}\right)\Delta_{\left|j\right)}\label{eq:tiga-1}\\
 & \quad\;\;\;+\,^{3}R_{\left(ij\right)}+\left(\mathrm{tr}\mathcal{K}+\Theta\left(\hat{n}\right)\right)K_{\left(ij\right)}-\mathcal{K}_{\left(i\right|}^{\,\:\:k}K_{k\left|j\right)}-K_{\left(i\right|k}\Delta_{\;\:\left|j\right)}^{k}=\kappa\mathcal{S}_{ij}-\frac{1}{2}\,^{3}q_{ij}\kappa\left(\mathcal{S}-E\right),\nonumber 
\end{align}
where we use the fact that $\mathcal{T=\mathcal{S}}-E$, with $\mathcal{S}=g^{ij}\mathcal{S}_{ij}$.
(\ref{eq:satu})-(\ref{eq:tiga-1}), are respectively, the energy,
momentum, and stress-energy equations for MAGR. Notice the existence
of the additional variables. One could show that by inserting the
Levi-Civita condition (\ref{eq:tor1}) and (\ref{eq:met1})-(\ref{eq:met}),
they return to the original (3+1)  Einstein field equation.

\section{Discussions and Conclusions}

\subsection{The Stress-Energy-Momentum Equation}

Equation (\ref{eq:satu})-(\ref{eq:tiga-1}) are the (3+1) decomposition
of the first Euler-Lagrange equation (\ref{eq:EoM1}); it comes from
the variation of action (\ref{eq:MAfR}) (for $f\left(\mathcal{R}\right)=\mathcal{R}$)
with respect to metric $g$. One could see that they provide 1+3+6=10
differential equations. For simplicity, let us take the time gauge
(or the Gauss normal coordinate \cite{Eric}), namely $N=1$, and
$\boldsymbol{N}=0$. Hence, equation (\ref{eq:satu})-(\ref{eq:tiga-1})
becomes:
\begin{align}
 & \frac{1}{2}\left(\,^{3}\mathcal{R}-\mathrm{tr}\left(K\mathcal{K}\right)+\left(\mathrm{tr}K\right)\left(\mathrm{tr}\mathcal{K}\right)-\left(\mathcal{K}_{i}^{\,\:j}+K_{\,\:i}^{j}\right)\Delta_{\;j}^{i}-\hat{n}\left[\mathrm{tr}\mathcal{K}\right]\right.\label{eq:satua}\\
 & \left.\quad+\,^{3}q^{ij}\hat{n}\left[K_{ij}\right]+\,^{3}\nabla_{i}\alpha^{i}-\,^{3}q^{ij}\,^{3}\nabla_{i}\Delta_{j}-\Theta_{i}\left(\alpha^{i}+\,^{3}q^{ij}\Delta_{j}\right)+\Theta\left(\hat{n}\right)\left(\mathrm{tr}\mathcal{K}+\mathrm{tr}K\right)\right)=\kappa E,\nonumber 
\end{align}
\begin{align}
 & \frac{1}{2}\left(\:\,^{3}\nabla_{j}\mathcal{K}_{i}^{\,\:j}-\,^{3}\nabla_{i}\mathcal{K}_{j}^{\,\:j}+\,^{3}T_{j\,\:i}^{\,\:\:k}\mathcal{K}_{k}^{\,\:j}-\alpha^{j}\left(K_{ij}+K_{ji}\right)\right.\label{eq:duaa}\\
 & \quad\;\;\;\left.+\,^{3}\nabla_{j}\Delta_{\;i}^{j}-\hat{n}\left[\,^{3}\omega_{j\,\:i}^{\,\:\:j}\right]+\left(\Delta_{i}+\Theta_{i}\right)\mathcal{K}_{j}^{\,\:j}+\left(\Delta_{j}-\Theta_{j}\right)\mathcal{K}_{i}^{\,\:j}-\partial_{i}\Theta\left(\hat{n}\right)+\hat{n}\left[\Theta_{i}\right]\right)=\kappa p_{i},\nonumber 
\end{align}
\begin{align}
\hat{n}\left[K_{\left(ij\right)}\right]-\Theta_{\left(i\right.}\Delta_{\left.j\right)}-\,^{3}\nabla_{\left(i\right.}\Delta_{\left.j\right)}+\,^{3}R_{\left(ij\right)}+\left(\mathrm{tr}\mathcal{K}+\Theta\left(\hat{n}\right)\right)K_{\left(ij\right)}-\mathcal{K}_{\left(i\right|}^{\,\:\:k}K_{k\left|j\right)}-K_{\left(i\right|k}\Delta_{\;\:\left|j\right)}^{k} & =\kappa\mathcal{S}_{ij}-\frac{1}{2}\,^{3}q_{ij}\kappa\left(\mathcal{S}-E\right),\label{eq:tigaa}
\end{align}
but the physical interpretation is invariant under the change of coordinate.

The energy equation (\ref{eq:satua}), contains time evolution (and
hence the dynamics) from the terms $\hat{n}\left[\mathrm{tr}\mathcal{K}\right]$
and $\hat{n}\left[K_{ij}\right]$. These originate from the term $\mathrm{\mathbf{Ric}}\left(\hat{n},\hat{n}\right)+\overline{\mathrm{\mathbf{Ric}}}\left(\hat{n},\hat{n}\right),$
which is not zero due to the non-metricity and torsion. The momentum
equation (\ref{eq:duaa}) also contains dynamics from $\hat{n}\left[\,^{3}\omega_{j\,\:i}^{\,\:\:j}\right]$
and $\hat{n}\left[\Theta_{i}\right]$. The first term contains the
change of 3-connection in time, while the second is the change of
the angle between $\boldsymbol{a}_{i}$ and $\hat{n}$. The second
term will vanish for the Levi-Civita connection. For the first term,
it enters the equation of motion because the connection is treated
as an independent variable in an equal footing with metric, in the
Palatini formulation. For the Levi-Civita case, one could show that
$\hat{n}\left[\,^{3}\omega_{j\:\,\:i}^{\:\:j}\right]=\,^{3}\nabla_{i}K_{j}^{\:\:j}.$
Notice also the explicit existence of the 3-torsion in the momentum
equation. The last equation, the stress-energy part, contains dynamics
via $\hat{n}\left[K_{\left[ij\right]}\right]$ as in the energy equation.
The torsion only enters the momentum equation explicitly, but it is
contained implicitly in all the equations, for example, in the antisymmetric
part of the extrinsic curvatures $K$ and $\mathcal{K}$. All these
three equations are dynamical.

One might ask where are the terms containing the time derivatives
of the 3-metric in equation (\ref{eq:satua})-(\ref{eq:tigaa}). In
the original EFE, equation (\ref{eq:satua}) and (\ref{eq:duaa})
become, respectively, the Hamiltonian and momentum (or diffeomorphism)
constraint with Levi-Civita condition (\ref{eq:tor1}) and (\ref{eq:met1})-(\ref{eq:met}),
there is no term containing the derivative with respect to time in
such equations because the additional parts cancel with each other:
\begin{align}
\frac{1}{2}\left(\,^{3}\mathcal{R}-\mathrm{tr}\left(K^{2}\right)+\left(\mathrm{tr}K\right)^{2}\right) & =\kappa E,\label{eq:nic1}\\
\,^{3}\nabla_{j}K_{i}^{\,\:j}-\,^{3}\nabla_{i}K_{j}^{\,\:j} & =\kappa p_{i}.\label{eq:nic2}
\end{align}
On the other hand, equation (\ref{eq:tigaa}), in the standard GR,
becomes:
\begin{equation}
\mathcal{L}_{\hat{n}}K_{ij}+\,^{3}R_{ij}+\left(\mathrm{tr}K\right)K_{ij}-2K_{i}^{\,\:\:k}K_{kj}=\kappa\mathcal{S}_{ij}-\frac{1}{2}\,^{3}q_{ij}\kappa\left(\mathcal{S}-E\right),\label{eq:nic}
\end{equation}
with $\mathcal{L}_{\hat{n}}K_{ij}$ is the Lie derivative of $K$
in the direction $\hat{n}$:
\[
\mathcal{L}_{\hat{n}}K_{ij}=\hat{n}\left[K_{ij}\right]+K_{ik}\partial_{j}n^{k}+K_{kj}\partial_{i}n^{k},
\]
where the last two terms in the RHS are zero in the normal coordinate.
The dynamics in (\ref{eq:tigaa}) is contained in the term $\hat{n}\left[K_{\left[ij\right]}\right]$,
however, the 3-metric and $K$ are related by the following equation:
\begin{equation}
\hat{n}\left[\,^{3}q_{ij}\right]=\left(\nabla_{\hat{n}}g\right)\left(\partial_{i},\partial_{j}\right)+K_{ij}+K_{ji},\label{eq:notnice}
\end{equation}
where for the Levi-Civita case becomes:
\begin{equation}
\hat{n}\left[\,^{3}q_{ij}\right]=2K_{ij},\label{eq:niceonemaan}
\end{equation}
hence, one could insert (\ref{eq:niceonemaan}) to (\ref{eq:nic})
to obtain the terms containing the double derivative of the 3-metric
with respect to the time coordinate $\partial_{0}^{2}\,^{3}q_{ij},$
giving the standard dynamics of GR. However, for the general affine
connection, $\nabla_{\hat{n}}g\neq0$, but: 
\[
\left(\nabla_{\hat{n}}g\right)\left(\partial_{i},\partial_{j}\right)=\hat{n}\left[\,^{3}q_{ij}\right]-\Delta_{ij}-\Delta_{ji},
\]
causing the $\hat{n}\left[\,^{3}q_{ij}\right]$'s in (\ref{eq:notnice})
to cancel each other, while leaving $K_{ij}+K_{ji}=\Delta_{ij}+\Delta_{ji},$
free from $\,^{3}q_{ij}$. Therefore, the momentum and stress-energy
equation (\ref{eq:duaa})-(\ref{eq:tigaa}) are free from the time
derivative of $\,^{3}q_{ij}$. However, the dynamics of the 3-metric
enters the energy equation (\ref{eq:satua}) from the term $\hat{n}\left[\mathrm{tr}\mathcal{K}\right]=\hat{n}\left[\,^{3}q_{ij}\mathcal{K}^{ij}\right]$.

One needs to keep in mind that at this stage, we are only working
with the first equation of motion (\ref{eq:EoM1}); there still exists
another equation of motion, namely, the one obtained from the variation
of the action with respect to the connection (\ref{eq:EoM2}). The
first equation of motion (\ref{eq:EoM1}) provides only 10 differential
equations, whereas the unknown variables are 74 (10 from the metric,
64 from the connections, assuming the theory does not have constraint).
With the additional variables on the hypersurface, we have introduced
64 unknown variables (1 for $\Theta\left(\hat{n}\right),$ 3 for each
$\Theta_{i}$, $\Delta_{i}$, $\,^{3}\alpha_{i}$, 9 for $K,$ $\mathcal{K}$,
$\Delta_{\;i}^{j}$, and 27 for $\,^{3}\omega$), and 10 more unknowns
should come from the metric $g_{\mu\nu},$ in terms of the 3-metric
$\,^{3}q_{ij},$ the lapse $N$, and the shift $\boldsymbol{N}.$
The second Euler-Lagrange equation (\ref{eq:EoM2}) will provide 60
more differential equations, leaving 4 vectorial degrees of freedom
on the connection. The last 4 equations come from the traceless torsion
constraint (\ref{eq:EoM3}), by taking the projective invariance transformation
(\ref{eq:projectivetranfs}) into account. Without the (3+1) decomposition
of these equations of motions as well, it is impossible to do a complete
analysis of the (3+1) MAGR theory.

\subsection{The Hypermomentum Equation and Traceless Torsion Constraint}

For the completeness of the discussion in this article, we add some
of the results from our companion paper. Here, we only present the
(3+1) decomposition of the hypermomentum equation in the normal (Gauss)
coordinate. The general treatment and the detailed derivation of the
result will be discussed in our companion paper.

\subsubsection*{(3+1) Decomposition of Hypermomentum}

The hypermomentum (\ref{eq:hypemom}) could be written with the spacetime
index $\alpha$ is hidden as $\mathcal{H}\left(\partial_{\mu},dx^{\beta}\right)=\mathcal{H}_{\:\:\mu}^{\alpha\,\:\beta}\partial_{\alpha}.$
Using this notation, one could decompose the hypermomentum by its
'internal' indices into the normal and parallel parts, with respect
to the hypersurface $\Sigma$:
\begin{align}
\mathcal{H}\left(\hat{n},\hat{n}^{*}\right) & =\kappa n^{\mu}n_{\beta}\mathcal{H}_{\:\:\mu}^{\alpha\,\:\beta}\partial_{\alpha}=-\left\langle \hat{n}^{*},\mathcal{H}\left(\hat{n},\hat{n}^{*}\right)\right\rangle \hat{n}+\left\langle \,^{3}dx^{i},\mathcal{H}\left(\hat{n},\hat{n}^{*}\right)\right\rangle \partial_{i},\nonumber \\
\mathcal{H}\left(\hat{n},\,^{3}dx^{i}\right) & =\kappa n^{\mu}\mathcal{H}_{\:\:\mu}^{\alpha\,\:i}\partial_{\alpha}=-\left\langle \hat{n}^{*},\mathcal{H}\left(\hat{n},\,^{3}dx^{i}\right)\right\rangle \hat{n}+\left\langle \,^{3}dx^{j},\mathcal{H}\left(\hat{n},\,^{3}dx^{i}\right)\right\rangle \partial_{j},\nonumber \\
\mathcal{H}\left(\partial_{i},\hat{n}^{*}\right) & =\kappa n_{\beta}\mathcal{H}_{\:\:i}^{\alpha\,\:\beta}\partial_{\alpha}=-\left\langle \hat{n}^{*},\mathcal{H}\left(\partial_{i},\hat{n}^{*}\right)\right\rangle \hat{n}+\left\langle \,^{3}dx^{j},\mathcal{H}\left(\partial_{i},\hat{n}^{*}\right)\right\rangle \partial_{j},\label{eq:h1}\\
\mathcal{H}\left(\partial_{i},\,^{3}dx^{j}\right) & =\kappa\mathcal{H}_{\:\:i}^{\alpha\,\:j}\partial_{\alpha}=-\left\langle \hat{n}^{*},\mathcal{H}\left(\partial_{i},\,^{3}dx^{j}\right)\right\rangle \hat{n}+\left\langle \,^{3}dx^{k},\mathcal{H}\left(\partial_{i},\,^{3}dx^{j}\right)\right\rangle \partial_{k}.\nonumber 
\end{align}
Following the decomposition of $\mathcal{T}$ into its 3 components,
i.e., the energy $E$, the momentum $p_{i}$, and the stress $\mathcal{S}_{ij}$,
one could apply the same procedure to the hypermomentum $\mathcal{H},$
where the decomposition is based on the split of the 'spacetime' into
space and time. This should not be confused with the split of $\mathcal{H}$
based on the symmetricity of the indices which gives the spin, shear,
and dilation parts as in \cite{Hehl}. 

\subsubsection*{The (3+1) Hypermomentum Equation in Normal Coordinate}

With the (3+1) decomposition in (\ref{eq:h1}), the hypermomentum
equation (\ref{eq:EoM2}) could be split into 4 equations as follows:
\begin{align}
 & n^{\mu}n^{\lambda}T_{\mu\;\;\;\lambda}^{\:\:\alpha}\partial_{\alpha}-n_{\beta}\left(\nabla_{\hat{n}}g^{\alpha\beta}\right)\partial_{\alpha}+n_{\beta}\left(\nabla_{\nu}g^{\nu\beta}\right)\hat{n}=\kappa n^{\mu}n_{\beta}\mathcal{H}_{\:\:\mu}^{\alpha\,\:\beta}\partial_{\alpha}-\frac{2}{3}n^{\mu}n_{\beta}\kappa\mathcal{H}_{\:\:\;\;\,\sigma}^{\left[\alpha\right|\,\:\sigma}\delta_{\mu}^{\left|\beta\right]}\partial_{\alpha},\label{eq:gokil}\\
 & n^{\mu}g^{i\lambda}T_{\mu\;\;\;\lambda}^{\:\:\alpha}\partial_{\alpha}+\frac{1}{2}\left(g_{\sigma\lambda}\nabla_{\nu}g^{\sigma\lambda}\right)\left(n^{\nu}g^{\alpha i}\partial_{\alpha}-g^{\nu i}\hat{n}\right)-\left(\nabla_{\hat{n}}g^{\alpha i}\right)\partial_{\alpha}+\left(\nabla_{\nu}g^{\nu i}\right)\hat{n}=\kappa n^{\mu}\mathcal{H}_{\:\:\mu}^{\alpha\,\:i}\partial_{\alpha}-n^{\mu}\delta_{\beta}^{i}\frac{2}{3}\kappa\mathcal{H}_{\:\:\;\;\,\sigma}^{\left[\alpha\right|\,\:\sigma}\delta_{\mu}^{\left|\beta\right]}\partial_{\alpha},\nonumber \\
 & n^{\lambda}T_{i\;\;\;\lambda}^{\:\:\alpha}\partial_{\alpha}+\frac{1}{2}\left(g_{\sigma\lambda}\nabla_{\nu}g^{\sigma\lambda}\right)\left(\delta_{i}^{\:\nu}\hat{n}-n^{\nu}\partial_{i}\right)-n_{\beta}\left(\nabla_{i}g^{\alpha\beta}\right)\partial_{\alpha}+n_{\beta}\left(\nabla_{\nu}g^{\nu\beta}\right)\partial_{i}=\kappa n_{\beta}\mathcal{H}_{\:\:i}^{\alpha\,\:\beta}\partial_{\alpha}-\delta_{i}^{\mu}n_{\beta}\frac{2}{3}\kappa\mathcal{H}_{\:\:\;\;\,\sigma}^{\left[\alpha\right|\,\:\sigma}\delta_{i}^{\left|\beta\right]}\partial_{\alpha},\nonumber \\
 & g^{j\lambda}T_{i\;\;\;\lambda}^{\:\:\alpha}\partial_{\alpha}+\frac{1}{2}\left(g_{\sigma\lambda}\nabla_{\nu}g^{\sigma\lambda}\right)\left(\delta_{i}^{\:\nu}g^{\alpha j}\partial_{\alpha}-g^{\nu j}\partial_{i}\right)-\left(\nabla_{i}g^{\alpha j}\right)\partial_{\alpha}+\left(\nabla_{\nu}g^{\nu j}\right)\partial_{i}=\kappa\mathcal{H}_{\:\:i}^{\alpha\,\:j}\partial_{\alpha}-\delta_{i}^{\mu}\delta_{\beta}^{j}\frac{2}{3}\kappa\mathcal{H}_{\:\:\;\;\,\sigma}^{\left[\alpha\right|\,\:\sigma}\delta_{\mu}^{\left|\beta\right]}\partial_{\alpha}.\nonumber 
\end{align}
Using equation (\ref{eq:h1}) together with torsion and non-metricity
decomposition in (\ref{eq:t1})-(\ref{eq:q2}), one could rewrite
the four equations (\ref{eq:gokil}) in terms of the additional variables
as follows:
\begin{align}
\left(\mathcal{K}_{i}^{\:\:i}-K_{i}^{\;i}\right)\hat{n}+\left(\Delta^{i}-\alpha^{i}\right)\partial_{i} & =\kappa\left(\mathcal{H}\left(\hat{n},\hat{n}^{*}\right)+\frac{1}{3}\left(\mathrm{tr}\mathcal{H}+\left\langle \hat{n},\mathrm{tr}\mathcal{H}\right\rangle \hat{n}\right)\right),\label{eq:hyp1}
\end{align}
\begin{align}
\left(\,^{3}q^{ij}\left(\Theta\left(\hat{n}\right)+\frac{1}{2}\,^{3}q_{kl}\left(n\left[\,^{3}q^{kl}\right]+\Delta^{kl}+\Delta^{lk}\right)\right)-\mathcal{K}^{ij}-n\left[\,^{3}q^{ij}\right]-\Delta^{ij}\right)\partial_{j}\label{eq:hyp2}\\
+\left(\Delta^{i}-2\Theta^{i}+\,^{3}\nabla_{j}\,^{3}q^{ji}-\frac{1}{2}\,^{3}q_{jk}\,^{3}\nabla^{i}\,^{3}q^{jk}\right)\hat{n} & =\kappa\left(\mathcal{H}\left(\hat{n},dx^{i}\right)+\frac{1}{3}\left\langle dx^{i},\mathrm{tr}\mathcal{H}\right\rangle \hat{n}\right),\nonumber 
\end{align}
\begin{align}
\left(\mathcal{K}_{j}^{\:\:j}-K_{j}^{\;j}+\Theta\left(\hat{n}\right)-\frac{1}{2}\,^{3}q_{kj}\left(n\left[\,^{3}q^{kj}\right]+\Delta^{kj}+\Delta^{jk}\right)\right)\partial_{i}\label{eq:hyp3}\\
+\left(K_{i}^{\;j}-\Delta_{\;\:i}^{j}\right)\partial_{j}+\left(\frac{1}{2}\,^{3}q_{jk}\,^{3}\nabla_{i}\,^{3}q^{jk}-\Delta_{i}\right)\hat{n} & =\kappa\left(\mathcal{H}\left(\partial_{i},\hat{n}^{*}\right)+\frac{1}{3}\left\langle \hat{n},\mathrm{tr}\mathcal{H}\right\rangle \partial_{i}\right),\nonumber 
\end{align}
\begin{align}
\left(\mathcal{K}_{i}^{\:\:j}-K_{\;i}^{j}\right)\hat{n}+\left(\Delta^{j}-\alpha^{j}-\Theta^{j}+\,^{3}\nabla_{k}\,^{3}q^{jk}-\frac{1}{2}\,^{3}q_{kl}\,^{3}\nabla^{j}\,^{3}q^{kl}\right)\partial_{i}\label{eq:hyp4}\\
+\left(\left(\Theta_{i}+\frac{1}{2}\,^{3}q_{lm}\,^{3}\nabla_{i}\,^{3}q^{lm}\right)\,^{3}q^{jk}-\,^{3}\nabla_{i}\,^{3}q^{jk}\right)\partial_{k}+\,^{3}q^{jk}\,^{3}T\left(\partial_{i},\partial_{k}\right) & =\kappa\left(\mathcal{H}\left(\partial_{i},dx^{j}\right)-\frac{1}{3}\left(\delta_{i}^{j}\mathrm{tr}\mathcal{H}-\left\langle dx^{j},\mathrm{tr}\mathcal{H}\right\rangle \partial_{i}\right)\right),\nonumber 
\end{align}
with $\mathrm{tr}\mathcal{H}=\mathcal{H}\left(\partial_{\mu},dx^{\mu}\right)=\mathcal{H}_{\:\:\mu}^{\alpha\,\:\mu}\partial_{\alpha}$.
These are the (3+1) hypermomentum equations in normal coordinates.
Notice that the quantity $\mathcal{H}\left(U,V^{*}\right)\in T_{p}\mathcal{M}$
is a vector and each equation still has the normal and parallel parts
with respect to $\Sigma$. Moreover, one could contract (\ref{eq:hyp1})-(\ref{eq:hyp4})
with $\hat{n}^{*}$ and $dx^{i}$ to obtain 8 equations.

\subsubsection*{The Traceless Torsion Constraint Decomposition}

As explained in the previous sections, the hypermomentum equation
(\ref{eq:EoM2}) only provides 60 equations; one needs 4 more equations
to reduce the vectorial degrees of freedom in the connection. These
are provided by the traceless torsion constraint (\ref{eq:EoM3}),
which could be decomposed (in normal coordinate) as follows. Notice
that $\boldsymbol{T}=T_{\nu}dx^{\nu}\in T_{p}^{*}\mathcal{M}$ is
a 1-form:
\[
T_{\nu}=T_{\mu\;\nu}^{\;\mu}=g\left(dx^{\mu},T\left(\partial_{\mu},\partial_{\nu}\right)\right)=0.
\]
By contracting $\boldsymbol{T}$ with the normal $\hat{n}$ and $\partial_{i}$,
then using torsion decomposition (\ref{eq:t1})-(\ref{eq:t3}), one
obtains (in normal coordinate):
\begin{align}
 & \left\langle \boldsymbol{T},\hat{n}\right\rangle =n^{\nu}T_{\nu}=n^{\nu}T_{\mu\;\nu}^{\;\mu}=\Delta_{\;\:i}^{i}-\mathcal{K}_{i}^{\:\:i}=0,\label{eq:proj1}\\
 & \left\langle \boldsymbol{T},\partial_{i}\right\rangle =T_{i}=T_{\mu\;i}^{\;\mu}=\Delta_{i}-\Theta_{i}+\,^{3}T_{j\;i}^{\;j}=0.\label{eq:proj2}
\end{align}

\subsubsection*{The Zero Hypermomentum Case}

When the hypermomentum is zero, the hypermomentum equation (\ref{eq:EoM2}),
together with the traceless torsion constraint (\ref{eq:EoM3}) must
give the condition for the Levi-Civita connection. Let us check if
the (3+1) equations (\ref{eq:hyp1})-(\ref{eq:hyp4}) agree with this
fact. Setting $\mathcal{H}=0$ and then contracting (\ref{eq:hyp1})-(\ref{eq:hyp4})
with $\hat{n}^{*}$ and $dx^{i}$, we obtain 8 equations, which include
1 scalar equation:
\begin{equation}
\mathcal{K}_{i}^{\:\:i}-K_{i}^{\;i}=0,\label{eq:sclr}
\end{equation}
3 vector equations:
\begin{align}
 & \Delta^{i}-\alpha^{i}=0,\label{eq:vect1}\\
 & \Delta^{i}-2\Theta^{i}+\,^{3}\nabla_{j}\,^{3}q^{ji}-\frac{1}{2}\,^{3}q_{jk}\,^{3}\nabla^{i}\,^{3}q^{jk}=0,\label{eq:vect2}\\
 & \frac{1}{2}\,^{3}q_{jk}\,^{3}\nabla_{i}\,^{3}q^{jk}-\Delta_{i}=0,\label{eq:vect3}
\end{align}
3 matrix equations:
\begin{align}
 & \,^{3}q^{ij}\left(\Theta\left(\hat{n}\right)+\frac{1}{2}\,^{3}q_{kl}\left(n\left[\,^{3}q^{kl}\right]+\Delta^{kl}+\Delta^{lk}\right)\right)-\mathcal{K}^{ij}-n\left[\,^{3}q^{ij}\right]-\Delta^{ij}=0,\label{eq:mat1}\\
 & K_{i}^{\;j}-\Delta_{\;\:i}^{j}+\left(\mathcal{K}_{k}^{\:\:k}-K_{k}^{\;k}+\Theta\left(\hat{n}\right)-\frac{1}{2}\,^{3}q_{kl}\left(n\left[\,^{3}q^{kl}\right]+\Delta^{kl}+\Delta^{lk}\right)\right)\delta_{i}^{j}=0,\label{eq:mat2}\\
 & \mathcal{K}_{i}^{\:\:j}-K_{\;i}^{j}=0,\label{eq:mat3}
\end{align}
and 1 tensor equation of order $\tbinom{2}{1}$:
\begin{align}
\left(\Delta^{j}-\alpha^{j}-\Theta^{j}+\,^{3}\nabla_{l}\,^{3}q^{jl}-\frac{1}{2}\,^{3}q_{lm}\,^{3}\nabla^{j}\,^{3}q^{lm}\right)\delta_{i}^{k}+\,^{3}q^{jl}\,^{3}T_{i\;\:l}^{\:k}\label{eq:tensor}\\
+\left(\Theta_{i}+\frac{1}{2}\,^{3}q_{lm}\,^{3}\nabla_{i}\,^{3}q^{lm}\right)\,^{3}q^{jk}-\,^{3}\nabla_{i}\,^{3}q^{jk} & =0.\nonumber 
\end{align}
The traceless torsion constraint (\ref{eq:EoM3}) provides 1 scalar
equation (\ref{eq:proj1}) and 1 vector equation (\ref{eq:proj2}).

These equations could be simplified as follows. Solving the vector
equations (\ref{eq:vect2})-(\ref{eq:vect3}) for $\Theta^{i}$ and
$\Delta^{i}$ gives:
\begin{align}
\Theta^{i} & =\frac{1}{2}\,^{3}\nabla_{j}\,^{3}q^{ji},\label{eq:vect2a}\\
\Delta^{i} & =\frac{1}{2}\,^{3}q_{jk}\,^{3}\nabla^{i}\,^{3}q^{jk},\label{eq:vect3a}
\end{align}
while solving the matrix equations (\ref{eq:mat1})-(\ref{eq:mat2})
gives:
\begin{align}
 & \Theta\left(\hat{n}\right)=\frac{1}{6}\,^{3}q_{kl}\nabla_{n}\,^{3}q^{kl},\label{eq:mata}\\
 & K^{ij}-\mathcal{K}^{ij}=\nabla_{n}\,^{3}q^{ij}-\frac{1}{3}\,^{3}q^{ij}\,^{3}q_{kl}\nabla_{n}\,^{3}q^{kl},\label{eq:matb}
\end{align}
with the help of the scalar equation (\ref{eq:sclr}). At last, using
the vector equations (\ref{eq:vect1}), (\ref{eq:vect2a}), and (\ref{eq:vect3a})
to (\ref{eq:tensor}), then decomposing this tensor equation into
the symmetric and antisymmetric parts of the $\left(i,j\right)$-indices,
gives:
\begin{align}
 & \,^{3}\nabla^{\left(i\right.}\,^{3}q^{\left.j\right)k}-\,^{3}q^{k\left(i\right.}\,^{3}\nabla_{l}\,^{3}q^{\left.j\right)l}=0,\label{eq:tensorsym}\\
 & \,^{3}T^{ikj}=\,^{3}q^{k\left[i\right.}\,^{3}q_{lm}\,^{3}\nabla^{\left.j\right]}\,^{3}q^{lm}+\,^{3}\nabla^{\left[i\right.}\,^{3}q^{\left.j\right]k}.\label{eq:tensorantsym}
\end{align}

Let us solve these simplified equations. The easiest way is to start
from (\ref{eq:tensorsym}), which is satisfied if:
\begin{equation}
\,^{3}\nabla^{i}\,^{3}q^{jk}=0.\label{eq:1}
\end{equation}
Inserting (\ref{eq:1}) to (\ref{eq:tensorantsym}) gives:
\begin{equation}
\,^{3}T^{ikj}=0,\label{eq:tors1}
\end{equation}
while inserting (\ref{eq:1}) to vector equation (\ref{eq:vect2a})
gives:
\begin{equation}
\Theta^{i}=0.\label{eq:2-1}
\end{equation}
Inserting (\ref{eq:tors1}) to (\ref{eq:proj2}) gives:
\begin{equation}
\Delta_{i}-\Theta_{i}=0.\label{eq:tors2}
\end{equation}
Now let us focus on the matrix equation (\ref{eq:mat3}) and (\ref{eq:matb}).
Applying the symmetries decomposition to the indices $(i,j)$ gives
the following conditions:
\begin{align}
 & K^{\left(ij\right)}-\mathcal{K}^{\left(ij\right)}=0,\label{eq:p}\\
 & K^{\left[ij\right]}+\mathcal{K}^{\left[ij\right]}=0,\label{eq:q}\\
 & K^{\left(ij\right)}-\mathcal{K}^{\left(ij\right)}=\nabla_{n}\,^{3}q^{ij}-\frac{1}{3}\,^{3}q^{ij}\,^{3}q_{kl}\nabla_{n}\,^{3}q^{kl},\label{eq:r}\\
 & K^{\left[ij\right]}-\mathcal{K}^{\left[ij\right]}=0,\label{eq:s}
\end{align}
where (\ref{eq:p})-(\ref{eq:q}) comes from (\ref{eq:mat3}), while
(\ref{eq:r})-(\ref{eq:s}) comes from (\ref{eq:matb}). Comparing
(\ref{eq:q}) and (\ref{eq:s}), then $K^{\left[ij\right]}=\mathcal{K}^{\left[ij\right]}=0$,
which means that both the extrinsic curvatures do not possess antisymmetric
parts. On the other hand, from (\ref{eq:p}) and (\ref{eq:r}), we
obtain $\mathcal{K}^{\left(ij\right)}=K^{\left(ij\right)},$ that
are equivalent with the following conditions:
\begin{align}
K^{ij} & =\mathcal{K}^{ij},\label{eq:6-1}\\
K^{ij} & =K^{ji},\label{eq:7-1}
\end{align}
and:
\begin{align*}
\nabla_{n}\,^{3}q^{ij}-\frac{1}{3}\,^{3}q^{ij}\,^{3}q_{kl}\nabla_{n}\,^{3}q^{kl}=0,
\end{align*}
which is satisfied if:
\begin{equation}
\nabla_{n}\,^{3}q^{ij}=0.\label{eq:8-1}
\end{equation}
Inserting (\ref{eq:8-1}) to (\ref{eq:mata}) gives:
\begin{equation}
\Theta\left(\hat{n}\right)=0.\label{eq:9-1}
\end{equation}
Finally, using (\ref{eq:q3}) and inserting (\ref{eq:sclr}), (\ref{eq:8-1}),
and (\ref{eq:9-1}) to (\ref{eq:mat2}) gives:
\begin{equation}
\mathcal{K}_{i}^{\;j}-\Delta_{\;\:i}^{j}=0.\label{eq:10}
\end{equation}

The conditions obtained from solving the trivial hypermomentum equation
could be classified as follows: \{(\ref{eq:tors1}),(\ref{eq:tors2}),
(\ref{eq:7-1}), (\ref{eq:10})\} and \{(\ref{eq:vect1}), (\ref{eq:1}),
(\ref{eq:2-1}), (\ref{eq:6-1}), (\ref{eq:8-1}), (\ref{eq:9-1})\}
are, respectively, the torsionless and metric condition (\ref{eq:tor1})
and (\ref{eq:met1})-(\ref{eq:met}) in the normal (Gauss) coordinate
where $N=1$ and $\boldsymbol{N}=0$. Hence, for zero hypermomentum,
the affine connection becomes Levi-Civita, and the (3+1) stress-energy-momentum
equations for MAGR (\ref{eq:satua})-(\ref{eq:tigaa}) return to the
original EFE (\ref{eq:nic1})-(\ref{eq:nic}). Notice that without
the traceless torsion constraint (\ref{eq:proj1})-(\ref{eq:proj2}),
it is not possible to retrieve the metric compatibility and torsionless
condition,  in the absence of hypermomentum. A more detailed analysis
of the hypermomentum equation will be discussed in our companion article.

\subsection{Conclusions and Further Remarks}

In this subsection, we summarize the results achieved from our work.
First, we have derived the generalized Gauss-Codazzi-Mainardi equations
and torsion (3+1) decomposition, which are valid for manifold with
any dimension. This is done in Section II with equation (\ref{eq:e})
and (\ref{eq:torsion1})-(\ref{eq:torsion2}) (or equation (\ref{eq:gcm1})-(\ref{eq:gcm2})
and (\ref{eq:T1})-(\ref{eq:T2}), written in terms of components)
as the results. Second, we have shown that a metric connection $\nabla$
on $\left(\mathcal{M},g\right),$ induce a connection $\,^{3}\nabla$
on the hypersurface $\left(\Sigma,\,^{3}q\right),$ which also satisfies
the metric compatibility with respect to $\,^{3}q$. A similar statement
is also valid for a torsionless connection. These had been proven
in Section III. Third, having the previous results in hand, we have
performed the ADM formulation to the generalized EFE, $\mathbf{Ric}-\frac{1}{2}\boldsymbol{g}\mathcal{R}=\kappa\mathcal{\boldsymbol{T}},$
for Metric-Affine General Relativity, resulting in 3 parts of the
equation: the energy equation (\ref{eq:satu}), momentum equation
(\ref{eq:dua}), and stress-energy equation (\ref{eq:tiga-1}), with
the help of the additional variables on the hypersurface introduced
in Section IV. Unlike the standard GR case, the energy and momentum
part of the Einstein field equations are not constraint equations,
since all of the 3 equations contain the derivative of quantities
with respect to the time coordinate. We have also shown in the discussion
that by applying metric compatibility (\ref{eq:met1}) and torsionless
condition (\ref{eq:tor1}) to (\ref{eq:satu})-(\ref{eq:tiga-1}),
one will recover the Hamiltonian and momentum (or diffeomorphism)
constraint, together with the standard dynamics of GR. 

For the completeness of the discussion in this article, we present
some results of our companion paper, which is the (3+1) decomposition
(in normal coordinates) of the hypermomentum equation (\ref{eq:hyp1})-(\ref{eq:hyp4})
and the traceless torsion constraint (\ref{eq:proj1})-(\ref{eq:proj2}).
We have shown that in the (3+1) perspective, setting hypermomentum
to zero, one will recover the metric compatibility and torsionless
condition, hence, the connection must be Levi-Civita. These conditions,
in MAGR (and MAG) is not \textit{a priori} given but as a consequence
of the absence of hypermomentum.

However, the tracelessness of torsion in (\ref{eq:ELmagr31}) does
not result from the equation of motion, it is introduced in the kinematical
level, via the projective transformation. If one expects to derive
the tracelessness of torsion from the dynamics, the action must be
added with new terms at least quadratic in torsion \cite{Percacci}.
This will be an interesting subject to pursue.

Another remark is concerning the way to break the projective invariance.
As we had mentioned at the beginning of Section IV, fixing the torsion
to be traceless is only one of the possibilities to break the invariance.
It is also possible to choose the non-metricity factor, or moreover,
a linear combination of the non-metricity factor and torsion to be
traceless, as discussed in \cite{Iosifidis4,Smalley}. These choices
might result in different (3+1) decompositions, in particular, the
form of equation (\ref{eq:proj1})-(\ref{eq:proj2}), and will also
be an interesting subject for further works.

As we had already mentioned in the discussion, in this article, the
focus is on the first equation of motion (\ref{eq:EoM1}); we are
writing another article that focuses on the hypermomentum equation
and the projective invariant constraint, as a companion to this article.
With a complete ADM decomposition to these equations of motions as
well, it will be possible to do a complete analysis on the (3+1) formulation
of MAGR, in particular, the Cauchy problem and/or the Hamiltonian
analysis of the theory. As further works, it will be interesting to
perform the (3+1) ADM formulation to more general theories such as
Metric-Affine $f\left(\mathcal{R}\right)$-Gravity. We expect our
preliminary work in this article could motivate more research in this
direction.

\section*{Acknowledgement}

This research was funded by the P3MI-ITB and PDUPT DIKTI programs.

\end{document}